\documentclass[a4paper,11pt]{article}
\usepackage{jheppub} 
\usepackage{lineno}

\usepackage{orcidlink}

\usepackage[utf8]{inputenc}
\pdfoutput=1
\usepackage{graphicx}
\usepackage{amsmath}
\usepackage{amsfonts}
\usepackage{amssymb}
\usepackage{xcolor,soul}
\usepackage{epstopdf}
\usepackage{xcolor}
\usepackage{csquotes}

\usepackage{float}
\usepackage{subfigure}

\usepackage{relsize}
\usepackage{graphics}
\usepackage{hyperref}
\usepackage{mathrsfs}
\usepackage{amssymb}
\usepackage{booktabs}
\usepackage[normalem]{ulem}
\usepackage{amsthm}
\usepackage{cancel}
\usepackage{fontawesome}   
\usepackage{tabularx,ragged2e}

\usepackage{tikz}
\usetikzlibrary{positioning,decorations.pathmorphing}

\hypersetup{
     colorlinks   = true,
     citecolor    = blue,
     linkcolor    = blue,
     urlcolor     = blue,
}
\usepackage{physics}

\title{\boldmath  Evaporation of Primordial Black Holes in a Thermal Universe: A Thermofield Dynamics Approach}

\author[a]{Ayan Chatterjee\,\orcidlink{0000-0002-7461-6540},}
\author[b]{Jitumani Kalita\,\orcidlink{0009-0001-5681-6194}}
\author[b]{and Debaprasad Maity\,\orcidlink{0000-0002-5458-7121}}

\affiliation[a]{Department of Physics \& Astronomical Science, Central University of Himachal Pradesh, Dharamshala-
176215, India}
\affiliation[b]{Department of Physics, Indian Institute of Technology, Guwahati, 
Assam, India}

\emailAdd{ayan.theory@gmail.com}
\emailAdd{k.jitumani@iitg.ac.in}
\emailAdd{debu@iitg.ac.in}

\abstract{We investigate the impact of a finite temperature environment on the Hawking radiation from black holes (BHs), with particular focus on Kerr BHs immersed in a cosmological thermal bath. The emitted particles from BHs interact with the thermal background and thermalize, leading to a modification in the Hawking radiation spectrum. By employing the methods of Thermofield Dynamics (TFD), a real time formalism of thermal quantum field theory, we derive the modified occupation numbers of the Hawking spectrum for asymptotically flat spacetimes like the Schwarzschild and the Kerr geometries. These corrections depend on the interplay between the BH temperature and the ambient bath temperature. We apply this formalism in the early universe reheating background scenario arising after inflation and demonstrate that the thermal correction to Hawking spectrum enhances the evaporation rate of primordial black holes (PBHs). As a result, the lifetime of PBH shortens compared to the zero temperature vacuum and leads to interesting cosmological consequences.}


\newcommand{\bea}{\begin{aligned}}
\newcommand{\eea}{\end{aligned}}
\def\beq{\begin{equation}}
\def\eeq{\end{equation}}
\def\beqa{\begin{eqnarray}}
\def\eeqa{\end{eqnarray}}
\def\be{\begin{equation}}
\def\ee{\end{equation}}

\def\bse{\begin{subequations}}
\def\ese{\end{subequations}}

\def\bea{\begin{eqnarray}}
\def\eea{\end{eqnarray}}

\usepackage{pgfplots}
\pgfplotsset{compat=1.17}

\begin{document}

\maketitle

\flushbottom

\section{Introduction}

The primordial black holes (PBHs) are black holes (BHs) that may have formed in the early universe due to the collapse of large density fluctuations~\cite{Carr:1974, Young:2019yug, Jedamzik:1996mr, Kawaguchi:2007fz, Kim:1999xg, Yoo:2024lhp, Yoo:2022mzl, Escriva:2021aeh}, phase transitions~\cite{Jedamzik:1999am, Baker:2021nyl, Lewicki:2023ioy, Musco:2023dak, Flores:2024lng, Goncalves:2024vkj, Sobrinho:2016fay}, or other high-energy processes occurring before the Big Bang Nucleosynthesis (BBN)~\cite{Keith:2020jww, Conzinu:2023fui, Banerjee:2022xft, Green:2014faa, LISACosmologyWorkingGroup:2023njw, Kim:2024gqp}. Unlike stellar BHs, which result from the gravitational collapse of massive stars, PBHs can span a vast range of masses- from the Planck mass ($\sim 10^{-5}\, \rm g$) to several solar masses- depending on the formation epoch and mechanism~\cite{Frampton:2015png, Kirillov:2021qjz, Khlopov:2008qy, DeLuca:2020ioi, Wang:2025hwc, Saini:2017tsz}. One of the key signatures of PBHs is their evaporation \emph{via} Hawking radiation, a quantum process that leads to a slow loss of mass and eventual disappearance of BHs~\cite{Hawking:1975vcx}.

Hawking radiation, originally derived through a semiclassical framework for BHs in asymptotically flat, empty spacetimes, predicts that BHs emit particles with a thermal spectrum determined by their surface gravity. The conventional treatment assumes a vacuum background, wherein the dynamics of the quantum field modes are subject to the geometry of the black hole spacetime. However, in a realistic early Universe scenarios, PBHs are not isolated systems. They are surrounded by a hot, dense thermal bath comprised of particles in equilibrium at a finite temperature~\cite{Dimopoulos:2019wew, He:2022wwy, Hamaide:2023ayu}. The emitted particles from the BH can interact with the thermal bath and eventually thermalise which in turn may significantly alter the emission process. This expectation stems from the well known fact of 
analogous phenomena observed in early-universe, particularly during the decay of a field into radiation in a thermal environment. Even though the field itself may not be in thermal equilibrium with the background bath, the thermalisation of the produced particles leads to a modification of the decay width \cite{Garcia:2020wiy, Ahmed:2022tfm, Haque:2023yra, Adshead:2019uwj}. This occurs because finite-temperature quantum field theory modifies the effective phase space occupations of the final states through Bose enhancement and Pauli blocking factors.
By analogy, when a black hole is embedded in a thermal background, equilibration between the emitted Hawking flux and the thermal bath can lead to a similar modification of the emission spectrum. The modified Hawking flux we discuss is thus conceptually parallel to the thermal enhancement in particle decay rates observed in finite-temperature field theory.

To develop this notion of an altered Hawking spectrum due to ambient thermal environment, we shall employ Thermo Field Dynamics (TFD)~\cite{Takahashi:1996zn, Das_book, Matsumoto:1982ry, Mustafa:2022got, Nair:2015rsa, Umezawa1982, Ojima:1981ma}, a real time formalism of thermal quantum field theory. TFD allows one to represent thermal averages as expectation values in a doubled Hilbert space, enabling a direct operator-level treatment of thermal effects. In this work, we shall use TFD to reformulate the quantisation of the scalar and the Dirac field in the Schwarzschild and the Kerr BH backgrounds, in presence of a thermal bath. For the Schwarzschild geometry, the thermal correction modifies the occupation number of the outgoing modes, leading to an enhanced particle production that depends on both on the BH and the bath temperatures. For the Kerr spacetime, due to the presence of rotational Killing vectors, the energy of emitted particles depend both on the spin angular momentum of the BH as well as the background bath temperature. While earlier study~\cite{Kalita:2025foa} have explored thermal corrections for static Schwarzschild black holes, the present work extends this framework to the more general and astrophysically relevant Kerr geometry. Specifically, we derive the modified Hawking spectrum for both scalar and Dirac fields, highlighting how the interplay between the Hawking temperature, the ambient bath temperature, and the black hole spin affects the evaporation dynamics.

Beyond this formal derivation, we shall also explore the implications of our results in a cosmological context. In particular, we apply our formalism to the reheating phase that follows inflation. During reheating, the Universe transitions from a inflaton dominated phase to a radiation-dominated phase through the decay of the inflaton field, reponsible for inflation, into standard model particles~\cite{Kofman:1994rk, Kofman:1997yn, Bassett:2005xm, Allahverdi:2010xz, Moss:2008lkw, Shtanov:1994ce, Haque:2022kez, Haque:2020zco, Greene:1997fu, Amin:2014eta}. PBHs formed during this epoch would thus naturally be immersed in a thermal bath with time dependent temperature. We adopt a model independent parametrization of reheating dynamics to study how the thermal corrections to Hawking radiation affect the evolution of PBH. Our analysis reveals that the modified Hawking flux due to the thermal bath accelerates the evaporation of PBHs and leads to a reduction in PBH lifetime compared to the standard zero temperature scenario. This may have important consequences for the constraints on PBH abundance and other cosmological observations~\cite{Carr:2009jm, Carr:2020xqk, Oncins:2022ydg, Carr:2016drx, Green:2020jor, Das:2025vts, Irges:2025idm, Irges:2022oxk, Stojkovic:2004hz, Stojkovic:2005zh}.

The structure of this paper is as follows: Section~\ref{Zero_Hawking} provides a review of the standard derivation of Hawking radiation for both the Schwarzschild and the Kerr spacetimes. In Section~\ref{TFD}, we introduce the TFD formalism as a framework for analyzing quantum fields at finite temperature. We then apply this formalism in Section~\ref{Finite_Hawking} to compute the corrected Hawking spectrum for BHs immersed in a thermal bath. In Section~\ref{BH_decay}, we derive the resulting equations governing the evolution of the BH's mass and spin. The cosmological implications are explored in Section~\ref{Reheating}, where we study the evolution of PBHs within a model-independent reheating scenario. Finally, we summarize our findings and discuss future directions in Section~\ref{conclusion}.


\section{Hawking radiation at zero temperatures}\label{Zero_Hawking}
Hawking radiation is a quantum phenomenon through which BHs emit thermal radiation due to quantum field effects in curved spacetime. In the original derivation due to Hawking~\cite{Hawking:1975vcx}, the Hawking effect is assumed as a scattering process of quantum waves in the matter collapsing geometry, where black hole is the final state the collapse. The black body spectrum arises due to the altered structure of vacuum during the collapse process. Alternatively, the Hawking process can also be interpreted as a particle-antiparticle pair production near the event horizon, where one particle escapes to infinity while the other falls into the BH, leading to a net loss of mass and energy~\cite{Page:1976df}. The resulting radiation spectrum resembles that of a blackbody with a temperature proportional to the surface gravity of the BH. This mechanism implies that BHs are not entirely black but slowly evaporate over time. A host of methods have been developed for a deeper understanding of Hawking radiation, including the use of gauge and gravitational anomaly \cite{Robinson:2005pd, Iso:2006wa}, quantum tunneling \cite{Damour:1976jd, Parikh:1999mf}, and quasilocal methods \cite{Chatterjee:2012um}. All these methods have their own advantages and disadvantages. Here, we shall go through the standard calculation for Hawking radiation for asymptotically flat spacetimes like the Schwarzschild and the Kerr BHs \cite{Hawking:1975vcx}.

\subsection{Schwarzschild black holes}
We begin our analysis with the simplest BH solution in general relativity---the Schwarzschild BH. This solution, which describes a static, spherically symmetric, uncharged BH, can be express by the metric (we use the natural units with $c=1$ and $\hbar=1$)~\cite{1916SPAW.......189S}
\begin{equation}
ds^2 = -\left(1 - \frac{2GM}{r}\right) dt^2 + \left(1 - \frac{2GM}{r}\right)^{-1} dr^2 + r^2 (d\theta^2 + \sin^2\theta \, d\phi^2),
\end{equation}
where $M$ denotes the mass of the BH, and $G$ is the universal gravitational constant. This metric possesses a coordinate singularity at $r = 2GM$, which corresponds to the event horizon. To further the analysis near this horizon, it is useful to introduce the tortoise coordinate $r^{*}(r)$, defined through the following function:
\begin{equation}
r^* = r + 2GM\ln\left(\frac{r}{2GM} - 1\right).
\end{equation}
This transformation maps the semi-infinite region $r \in (2GM, \infty)$ to $r^* \in (-\infty, \infty)$ and removes the coordinate singularity from the metric in the radial time sector. The metric in $(t, r^*)$ coordinates becomes
\begin{equation}\label{Sch_metric}
ds^2 = \left(1 - \frac{2GM}{r}\right)(-dt^2 + dr^{*2}) + r^2 (d\theta^2 + \sin^2\theta \, d\phi^2).
\end{equation}
%

\subsubsection{Scalar fields}
We consider a massless scalar field $\Phi$ propagating in this BH background \eqref{Sch_metric}. The field obeys the Klein--Gordon equation,
\begin{equation}
\Box \Phi \equiv \frac{1}{\sqrt{-g}}\partial_{\mu}\left(\sqrt{-g}g^{\mu\nu}\partial_{\nu}\Phi \right) = 0.
\end{equation}
In the Schwarzschild background, the explicit form of the wave equation becomes
\begin{align}
-r^2\sin\theta\, \partial_t^2 \Phi + \sin\theta\, \partial_{r^{\ast}} \left(r^2 \partial_{r^{\ast}} \Phi \right) + \left(1 - \frac{2GM}{r} \right) \left\lbrace \partial_\theta(\sin\theta\, \partial_\theta \Phi) + \frac{1}{\sin\theta} \partial^2_\phi \Phi \right\rbrace = 0.
\end{align}
Solving this equation involves separating variables and analyzing the resulting radial and angular equations. We set the ansatzs as $ \Phi(t, r^*, \theta, \phi) = R(t, r^*) Y(\theta, \phi) $, which yields
\begin{align}
& \partial_{r^{\ast}} \left(r^2 \partial_{r^{\ast}} R \right)-r^2 \partial_t^2 R - \lambda_l \left(1 - \frac{2GM}{r} \right) R = 0, \label{R_T_eq} \\
& \frac{1}{\sin\theta} \partial_\theta(\sin\theta\, \partial_\theta Y) + \frac{1}{\sin^2\theta} \partial^2_\phi Y + \lambda_l Y = 0. \label{Y_eq}
\end{align}
Equation~\eqref{Y_eq} admits spherical harmonic solutions $ Y_{lm}(\theta,\phi) $ with eigenvalues $ \lambda_l = l(l+1) $. The radial wave equation, Eq.~\eqref{R_T_eq}, can be transformed into a more familiar Schr\"odinger-like form. This is accomplished by separating the harmonic time dependence and redefining the radial function $R(t, r^*)$ via the ansatz
\begin{equation}
    R(t, r^*) = \frac{1}{r} e^{-i\omega t} U(r^*) .
\end{equation}
This substitution yields a one-dimensional wave equation for the new function $U(r^*)$ as
\begin{equation}
    \frac{d^2 U}{d{r^*}^2} + \left(\omega^2 - V_{\rm eff}(r) \right) U = 0,
\end{equation}
where the effective potential, $V_{\rm eff}(r)$, is given by
\begin{equation}
    V_{\rm eff}(r) = \left(1 - \frac{2GM}{r} \right)\left( \frac{l(l+1)}{r^2} + \frac{2GM}{r^3} \right).
\end{equation}
In the asymptotic limits, both near the event horizon ($r \to 2GM$, corresponding to $r^* \to -\infty$) and at spatial infinity ($r \to \infty$, corresponding to $r^* \to \infty$), the effective potential vanishes ($V_{\rm eff} \to 0$). In these regions, the solutions for $U(r^*)$ become simple plane waves
\begin{equation}
    U(r^*) \sim e^{\pm i \omega r^*}.
\end{equation}
Reconstructing the full four-dimensional solution from these radial plane waves and the angular spherical harmonics, $Y_{lm}(\theta, \phi)$, gives the asymptotic form of the scalar field
\begin{equation}
    \Phi(t, r, \theta, \phi) \sim \frac{1}{r} e^{-i \omega t} e^{\pm i \omega r^*} Y_{lm}(\theta, \phi).
\end{equation}

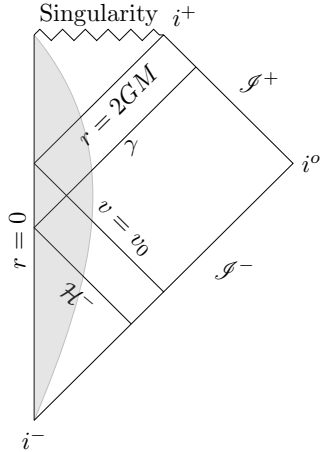
\begin{figure}[t]
\centering
\resizebox{0.3\linewidth}{!}{
\begin{tikzpicture}[node distance=2cm]
\coordinate (A) at (0,0);
\coordinate (B) at (2,-2);
\coordinate[label= above right:\normalsize{$i^{+}$}] (C) at (2,2);
\coordinate[label= right:\normalsize{$i^o$}] (D) at (4,0);
\coordinate[label= below:\normalsize{$i^{-}$}] (E) at (0,-4);
\coordinate (F) at (0,2);
\coordinate[label={[rotate=-40]\normalsize{$\mathcal{H}^-$}}] (G) at (0.5,-2.3);
\coordinate[below right = 0.707cm of C] (c);
\coordinate[below left = 0.707cm of B] (b);
\coordinate (a) at (0,-1);
\coordinate (f) at (0.4,2);

\draw (A) -- node[above,pos=0.6,sloped]{\normalsize $v=v_0$} (B) -- node[below right,pos=0.32] {\normalsize $\mathscr{I}^-$} (D) -- node[above right,pos=0.49] {\normalsize $\mathscr{I}^+$} (C) --node[below,sloped,pos=0.45] {\normalsize $r=2GM$} (A);
\draw (E) -- (B);
\draw (E) -- node[above,sloped,pos=0.7]{\normalsize $r=0$} (A);
\draw (A) -- (F);
\draw[decorate,decoration=zigzag] (F) -- node[above] {\normalsize Singularity} (C);
\draw (b) -- (a) -- node[below,pos=0.6]{\normalsize $\gamma$} (c);
\draw [fill=gray!80,nearly transparent] (E) to [out=67.5,in= -50.5] (F) ;
\end{tikzpicture}
}
\caption{Penrose diagram of a collapsing star. The light ray at $v=v_{0}$ is the last null ray scattered from the black hole geometry.}
\label{Penrose_diagram}
\end{figure}

In the context of BH radiation, we quantize the scalar field in the asymptotic regions: the past null infinity $\mathscr{I}^-$ and the future null infinity $\mathscr{I}^+$, as depicted in Fig.~\ref{Penrose_diagram}. These are the natural surfaces for defining in and out states in a scattering framework. This setup allows us to relate the in and out modes via Bogoliubov transformations and compute the particle content seen by an asymptotic observer. The mismatch between the vacua associated with $\mathscr{I}^-$ and $\mathscr{I}^+$ results in particle creation, which is the essence of Hawking radiation.

In the past null infinity $\mathscr{I}^-$, we can decomposed the field as
\begin{equation}
\hat{\Phi}=\int_0^{\infty} d\omega \sum_{l,m} \left(f_{\omega l m}\hat{a}^-_{\omega lm}+f^{\ast}_{\omega lm}\hat{a}^+_{\omega lm} \right),
\end{equation}
where, $\lbrace f_{\omega l m}\rbrace $ are the ingoing modes and are positive frequency with respect to some affine parameter define on $\mathscr{I}^-$
\begin{equation}\label{f_form_sch}
f_{\omega l m}=\frac{1}{\sqrt{2\pi \omega}}\frac{1}{r}e^{-i\omega v}Y_{lm}(\theta,\phi),
\end{equation}
with $v=(t+r^*)$ is the tortoise ingoing null coordinate and $\hat{a}^-_{\omega l m}$ and $\hat{a}^+_{\omega l m}$ are the annihilation and creation operators respectively. The vacuum $\vert 0_-\rangle $ on $\mathscr{I}^-$ is defined as $\hat{a}^-_{\omega l m} \vert 0_-\rangle=0$. Those modes $\lbrace f_{\omega l m}\rbrace $ forms complete orthonormal sets on $\mathscr{I}^-$ as 
\begin{align}\label{normal_eq}
\left(f_{\omega l m},f_{\omega^{\prime} l^{\prime} m^{\prime}} \right)_{\mathscr{I}^-} &=\delta(\omega-\omega^{\prime})\delta_{l l^{\prime}} \delta_{m m^{\prime}} , \;\;\;\; \left(f^{\ast}_{\omega l m},f^{\ast}_{\omega^{\prime} l^{\prime} m^{\prime}} \right)_{\mathscr{I}^-} =-\delta(\omega-\omega^{\prime})\delta_{l l^{\prime}} \delta_{m m^{\prime}}, \notag \\
\left(f_{\omega l m},f^{\ast}_{\omega^{\prime} l^{\prime} m^{\prime}} \right)_{\mathscr{I}^-} &=0.
\end{align}
The inner product, defined on the space of solutions of the KG equation, is defined as~\cite{Birrell:1982ix}
\begin{equation} \label{inner}
\left(A,B \right)_{\Sigma}=-\frac{i}{2}\int_{\Sigma}\left(A\partial_{\mu} B^{\ast}-B^{\ast}\partial_{\mu} A \right)\,\sqrt{-g}\, d\Sigma^{\mu},
\end{equation}
where $A, B$ are solutions of the KG equation, $d\Sigma^{\mu}=n^{\mu}d\Sigma$, with $n^{\mu}$ a future directed unit vector normal to the hypersurface $\Sigma$ and $d\Sigma$ is the volume element on $\Sigma$. 

A similar decomposition must be performed for the quantum field on future null infinity, $\mathscr{I}^+$. A complete basis in this region must account for modes that propagate outwards to future observers (the `outgoing' modes) as well as modes that are lost behind the event horizon (the `ingoing' modes). The field operator $\hat{\Phi}$ is therefore expanded on this complete basis as
\begin{equation}
\hat{\Phi}=\int_0^{\infty} d\omega \sum_{l , m} \left(p_{\omega l m}\hat{b}^-_{\omega l m}+p^{\ast}_{\omega l m}\hat{b}^+_{\omega l m}+q_{\omega l m}\hat{c}^-_{\omega l m}+q^{\ast}_{\omega l m}\hat{c}^+_{\omega l m} \right),
\end{equation}
where, $\lbrace p_{\omega l m}\rbrace$ and $\lbrace q_{\omega l m}\rbrace$ are the positive frequency outgoing and ingoing modes respectively, given by
\begin{equation}\label{p_form_sch}
p_{\omega l m}=\frac{1}{\sqrt{2\pi \omega}}\frac{1}{r}e^{-i\omega u}Y_{lm}(\theta,\phi), \quad q_{\omega l m}=\frac{1}{\sqrt{2\pi \omega}}\frac{1}{r}e^{-i\omega v}Y_{lm}(\theta,\phi).
\end{equation}
Here $u=(t-r^*)$ and $v=(t+r^*)$ are the tortoise null coordinates; and $\hat{b}^-_{\omega l m}$ and $\hat{b}^+_{\omega l m}$ are the annihilation and creation operators respectively for outgoing modes. Those modes $\lbrace p_{\omega l m}\rbrace$ and $\lbrace q_{\omega l m}\rbrace$ satisfies the orthonormality condition (\ref{normal_eq}) on future null infinity $\mathscr{I}^+$ and past event horizon $\mathcal{H}^-$ respectively. The vacuum $\vert 0_+\rangle $ on $\mathscr{I}^+$ is defined via $\hat{b}^-_{\omega l m} \vert 0_+\rangle=0$ or $\hat{c}^-_{\omega l m} \vert 0_+\rangle=0$.

In the subsequent sections, we proceed to derive the Hawking spectrum explicitly for this configuration using the standard method of mode scattering using the Bogoliubov coefficients relating the modes at different vacuum states.

\subsubsection*{Particle Creation}
To understand Hawking radiation as a particle creation phenomenon, we analyze how wave modes propagate through the BH geometry following \cite{Hawking:1975vcx, Parker:2009uva}. We focus on a mode $p_{\omega lm}$ defined on future null infinity $\mathscr{I}^+$ and trace it backward in time along a null trajectory $\gamma$, as illustrated in Fig.~\ref{Penrose_diagram}. As it propagates, part of the mode enters the BH horizon (denoted $p_{\omega lm}^{(1)}$), while the remainder is scattered back to past null infinity $\mathscr{I}^-$ (denoted $p_{\omega lm}^{(2)}$). These components are orthogonal due to their distinct causal structures, allowing the mode to be expressed as a linear combination~\cite{Hawking:1975vcx}
\begin{equation}
p_{\omega lm} = p_{\omega lm}^{(1)} + p_{\omega lm}^{(2)}.
\end{equation}
The normalization conditions of these parts are given by their  corresponding inner products,
\begin{equation}
\begin{aligned}
\left( p_{\omega l m}^{(2)}, p_{\omega^{\prime}l^{\prime} m^{\prime}}^{(2)}\right) &= \Gamma_{\omega l m}\delta(\omega-\omega^{\prime}) \delta_{l l^{\prime}} \delta_{m m^{\prime}}, \\
\left( p_{\omega l m}^{(1)}, p_{\omega^{\prime}l^{\prime} m^{\prime}}^{(1)}\right) &= (1 - \Gamma_{\omega l m}) \delta(\omega-\omega^{\prime}) \delta_{l l^{\prime}} \delta_{m m^{\prime}},
\end{aligned}
\end{equation}
where $\Gamma_{\omega lm}$ is the greybody factor, representing the transmission probability for the mode to reach $\mathscr{I}^-$. The scattered component $p_{\omega lm}^{(2)}$ at $\mathscr{I}^-$ has the asymptotic form
\begin{equation}
p_{\omega lm}^{(2)} \approx \frac{1}{\sqrt{2\pi \omega}}\frac{1}{r}e^{-i\omega u(v)} Y_{lm}(\theta,\phi),
\end{equation}
where the coordinate transformation relating advanced and retarded null coordinates is derived by solving the Killing equation along the null geodesic $\gamma$ as~\cite{Hawking:1975vcx}
\begin{equation}\label{u_v_rel_sch}
u(v) = -\frac{1}{\kappa} \ln\left(\frac{v_0 - v}{K}\right),
\end{equation}
and $\kappa = 1/4GM$ is the surface gravity of the Schwarzschild BH, $K$ is some constant, and $v_0$ is the reference time. The scattered modes \( \{ p_{\omega l m}^{(2)} \} \) can be decomposed in terms of the complete set of modes \( \{ f_{\omega l m}, f^*_{\omega l m} \} \) defined on \( \mathscr{I}^- \), as
\begin{equation}
p_{\omega l m}^{(2)} =\int_0^{\infty} \hspace{-0.25cm} d\omega^{\prime} \sum_{l^{\prime}, m^{\prime}} \left(\alpha_{\omega l m \; \omega^{\prime} l^{\prime} m^{\prime}}f_{\omega^{\prime}l^{\prime} m^{\prime}}+\beta_{\omega lm \; \omega^{\prime}l^{\prime} m^{\prime}}f^{\ast}_{\omega^{\prime}l^{\prime} m^{\prime}} \right),
\end{equation}
where the Bogoliubov coefficients $\alpha_{\omega lm \; \omega^{\prime}l^{\prime} m^{\prime}}$ and $\beta_{\omega lm \; \omega^{\prime}l^{\prime} m^{\prime}}$ can be written using the orthonormality conditions of $\lbrace f_{\omega lm}\rbrace$, as
\begin{align}
\alpha_{\omega lm \; \omega^{\prime}l^{\prime} m^{\prime}} =\left(p_{\omega lm}^{(2)},f_{\omega^{\prime}l^{\prime} m^{\prime}} \right)_{\mathscr{I}^-},\;\;\;\;\; \beta_{\omega lm \; \omega^{\prime}l^{\prime} m^{\prime}} =-\left(p_{\omega lm}^{(2)},f^{\ast}_{\omega^{\prime}l^{\prime} m^{\prime}} \right)_{\mathscr{I}^-}, \label{alpha_beta}
\end{align}
with the normalization condition~\cite{Traschen:1999zr, Mukhanov:2007zz}
\begin{equation}\label{normal_alpha_beta}
\int_0^\infty d\omega' \sum_{l', m'} \left( |\alpha_{\omega lm \; \omega' l' m'}|^2 - |\beta_{\omega lm \; \omega' l' m'}|^2 \right) = \Gamma_{\omega lm} \delta(0).
\end{equation}
The initial quantum state $|0_-\rangle$, defined as the vacuum on $\mathscr{I}^-$, contains no incoming particles, i.e., $\hat{a}_{\omega lm}^- |0-\rangle = 0$. However, due to the non-trivial Bogoliubov mixing, an observer at $\mathscr{I}^+$ detects particles in this vacuum. The number operator expectation value is
\begin{equation}
\langle 0_- | \hat{b}_{\omega lm}^{+(2)} \hat{b}_{\omega lm}^{-(2)} | 0_- \rangle = \int_0^\infty d\omega' \sum_{l',m'} |\beta_{\omega lm \; \omega' l' m'}|^2,
\end{equation}
where $\hat{b}_{\omega lm}^{-(2)} = (p_{\omega lm}^{(2)}, \hat{\phi})$ and $\hat{b}_{\omega lm}^{+(2)} = (p_{\omega lm}^{(2)*}, \hat{\phi})$ denote the annihilation and creation operators for the scattered modes computed by utilizing the inner products Eq.~\eqref{inner}. Using Eq.~(\ref{alpha_beta}), the Bogoliubov coefficients can be calculated in the eikonal approximation ($\omega^{\prime} \ggg \kappa$) as
\begin{equation}\label{bogo_sch_boson}
\begin{aligned}
& \alpha_{\omega lm \; \omega' l' m'} \approx \frac{1}{4 \pi} \sqrt{\frac{\omega^{\prime}}{\omega}}e^{i\omega^{\prime} v_0} K^{-\frac{i\omega}{\kappa}} (i\omega^{\prime}) ^{-\left(\frac{i\omega}{\kappa}+1 \right)} \Gamma (\frac{i\omega}{\kappa}+1) \delta_{l l'} \delta_{m m'}, \\
& \beta_{\omega lm \; \omega' l' m'} \approx -i \alpha_{\omega lm \; (-\omega') l' m'},
\end{aligned}
\end{equation}
where $\Gamma(x)$ is the Gamma function. Using this relation, the Bogoliubov coefficients can be related through
\begin{equation}\label{alpha_beta_relation}
|\beta_{\omega lm \; \omega' l' m'}|^2 = e^{-\frac{2\pi \omega}{\kappa}} |\alpha_{\omega lm \; \omega' l' m'}|^2.
\end{equation}
Substituting into the normalization condition, we find the number of particles produced in the mode $(\omega, l, m)$ to be
\begin{equation}\label{sch_spectrum}
\langle 0_- | \hat{b}_{\omega lm}^{+(2)} \hat{b}_{\omega lm}^{-(2)} | 0_- \rangle = \frac{\Gamma_{\omega lm}}{e^{\frac{2\pi \omega}{\kappa}} - 1} \delta(0).
\end{equation}
Finally, summing over angular momentum modes, the number density of emitted particles per frequency mode becomes
\begin{equation}
n_{\omega} = \frac{\Gamma_{\omega}}{e^{\omega / T_{\rm BH}} - 1},
\end{equation}
where $\Gamma_{\omega} = \sum_{l,m} \Gamma_{\omega lm}$ is the total greybody factor and $T_{\rm BH} = \kappa / 2\pi = 1/8\pi GM$ is the Hawking temperature of the Schwarzschild BH.

\subsubsection{Fermionic fields} \label{sch_fermion_section}
We now turn our attention to the dynamics of spin-$\frac{1}{2}$ fermionic field, $\Psi$, within the Schwarzschild spacetime Eq.~\eqref{Sch_metric}. In curved spacetime, the evolution of a spinor field is governed by the covariant form of the Dirac equation. For a massless spinor, the equation reads
\begin{equation}\label{Dirac_eq}
 i\gamma^{\mu}\nabla_{\mu} \Psi=0,
\end{equation}
where $ \gamma^{\mu} $ are the spacetime dependent Dirac matrices, and $ \nabla_{\mu} = \partial_{\mu}+\Omega_{\mu} $ is the spinor covariant derivative. The spin connection $ \Omega_{\mu} $ is given by $ \Omega_{\mu} = \omega_{ab\, \mu}[\gamma^a, \gamma^b]/8 $, with $\omega_{ab\, \mu} = \eta_{ac}\left[e^c_{\nu}\partial_{\mu}e^{\nu}_b+ e^c_{\nu}e^{\sigma}_b\Gamma^{\nu}_{\sigma\mu} \right]$, and $\Gamma^{\nu}_{\sigma\mu} = \frac{1}{2}g^{\nu\alpha} \left[g_{\alpha\sigma,\mu}+g_{\alpha\mu,\sigma}-g_{\sigma\mu,\alpha}  \right]$ is the Christoffel symbol. The Dirac matrices $ \gamma^{\mu} $ satisfies the well-known Clifford algebra $ \{ \gamma^{\mu} , \gamma^{\nu}  \} = 2g^{\mu \nu}I_4 $, with $ I_4 $ being the identity operator and $ g^{\mu \nu} $ is the background spacetime metric.\\

The vector fields $e^{\mu}_a$ or $e_{\mu}^a$ called the tetrad components with the property of $e^{\mu}_a e^b_{\mu}=\delta_a^b$ or $e^{\mu}_a e^a_{\nu} = \delta^{\mu}_{\nu}$ are used to transform the gamma matrices from Minkowski background $\eta^{a b}$ to Schwarzschild spactime $g^{\mu \nu}$ as $\gamma^{\mu}=e^{\mu}_a \gamma^{a}$. Here we use the Greek indices $\mu,\nu~(=t,r^{\ast},\theta,\phi)$ for curved background and Latin indices $a,b~(=0,1,2,3)$ for Minkowski. The Dirac equation in the Schwarzschild background becomes~\cite{deOliveira:2019hty}
\begin{equation}\label{Dirac_eq_2}
i\partial_t \Psi-i\gamma^0 \Big[ \gamma^1 \left(\frac{1}{4}\frac{f^{\prime}}{f}+\partial_{r^{\ast}}+\frac{f}{r}\right)+\gamma^2 \frac{f^{\frac{1}{2}}}{r} \left( \partial_{\theta}+\frac{\cot\theta}{2}\right) +\gamma^3\frac{f^{\frac{1}{2}}}{r\sin\theta}\partial_{\phi}\Big] \Psi=0,
\end{equation}
where $f=1-2GM/r$ and $f^{\prime}=df/dr^*$.\\

Now we choose a specific representation for the gamma matrices, $\gamma^a$, in flat spacetime. Two commonly used representations are the diagonal (Weyl) gauge and the Cartesian (Dirac) gauge. In the Weyl representation, the chirality operator is diagonal, given by $\gamma^5 = i\gamma^0\gamma^1\gamma^2\gamma^3 = \mathrm{diag}(-I_2, I_2)$. This form is particularly useful because it explicitly decouples the Dirac equation into two independent two-component Weyl equations.
In contrast, the Cartesian representation preserves manifest covariance under spatial rotations ($SO(3)$), making it especially suitable for solving the angular part of the Dirac equation using spinor spherical harmonics~\cite{Villalba:1995pg, Cotescu1997}.
The diagonal gauge has the following representation for the gamma matrices:
\begin{align}
\gamma^0 &= \gamma^0_d, \quad 
\gamma^1 = \gamma^1_d, \quad 
\gamma^2 = \gamma^2_d , \quad 
\gamma^3 = \gamma^3_d , 
\end{align}
and whereas, in the Cartesian gauge, we have,
\begin{equation}
\begin{aligned}
\gamma^0 &= \gamma^0_c = \gamma^0_d, \hspace{0.8cm}
\gamma^1 = \gamma^1_c = \left(\gamma^1_d \cos\phi + \gamma^2_d \sin\phi\right)\sin\theta + \gamma^3_d  \cos\theta,  \\
\gamma^2 &= \gamma^2_c = \left(\gamma^1_d \cos\phi + \gamma^2_d \sin\phi\right)\cos\theta - \gamma^3_d  \sin\theta,  \hspace{0.7cm}
\gamma^3 = \gamma^3_c = -\gamma^1_d \sin\phi + \gamma^2_d \cos\phi,
\end{aligned}
\end{equation}
where $\gamma^0_d=\begin{pmatrix}
i I_2 & 0 \\
0 & -i I_2
\end{pmatrix}$ and $\gamma^i_d =\begin{pmatrix}
0 & i \sigma^i \\
-i \sigma^i & 0
\end{pmatrix}$ and $\sigma^i$ are the Pauli matrices. See appendix~\ref{appendix_dirac_formalism} for detailed calculations. Those two sets of gamma matrices are related as $\gamma_c^a=S\gamma^a_d S^{-1}$, where the form of the matrix $S$ is
\begin{equation}
S=e^{-\frac{1}{2}\phi \gamma_d^1 \gamma_d^2}e^{-\frac{1}{2}\theta \gamma_d^3 \gamma_d^1}\frac{1}{2}\left(I_4 - \gamma_d^2 \gamma_d^1 -  \gamma_d^1 \gamma_d^3- \gamma_d^3 \gamma_d^2  \right).
\end{equation}
Now in the diagonal gauge the Dirac equation Eq.~\eqref{Dirac_eq_2} can be written as
\begin{equation}\label{dirac_diag}
i\partial_t \Psi_d - i \gamma^0_d\Big[\gamma^1_d \left(\frac{1}{4}\frac{f^{\prime}}{f}+\partial_{r^{\ast}}+\frac{f}{r}\right)+\gamma^2_d \frac{f^{\frac{1}{2}}}{r} \left( \partial_{\theta}+\frac{\cot\theta}{2}\right) +\gamma^3_d\frac{f^{\frac{1}{2}}}{r\sin\theta}\partial_{\phi}\Big] \Psi_d=0.
\end{equation}
Similarly in the Cartesian gauge the equation becomes
\begin{equation}
i\partial_t \Psi_c - i \gamma^0_c\Big[\gamma^1_c \left(\frac{1}{4}\frac{f^{\prime}}{f}+\partial_{r^{\ast}}+\frac{f}{r}\right)+\gamma^2_c \frac{f^{\frac{1}{2}}}{r} \left( \partial_{\theta}+\frac{\cot\theta}{2}\right) +\gamma^3_c\frac{f^{\frac{1}{2}}}{r\sin\theta}\partial_{\phi}\Big] \Psi_c=0  .
\end{equation}
The field in both the gauges can be related using the matrix $S$ as $\Psi_c=S \Psi_d$. We now rewrite the Dirac equation in the diagonal gauge, Eq.~\eqref{dirac_diag}, as an eigenvalue equation of the form $i\partial_t \Psi_d=H_d \Psi_d$, where the Hamiltonian $H_d$ is given by
\begin{equation}
H_d=\gamma^0_d\Big[\gamma^1_d \left(\frac{1}{4}\frac{f^{\prime}}{f}+\partial_{r^{\ast}}+\frac{f}{r}\right)+\gamma^2_d \frac{f^{\frac{1}{2}}}{r} \left( \partial_{\theta}+\frac{\cot\theta}{2}\right) +\gamma^3_d\frac{f^{\frac{1}{2}}}{r\sin\theta}\partial_{\phi}\Big] .
\end{equation}
To simplify this, we perform a unitary transformation using the operator $ \tilde{U}_d = \frac{1}{2} \Big(I_4  - \gamma_d^2 \gamma_d^1 \\-  \gamma_d^1 \gamma_d^3 - \gamma_d^3 \gamma_d^2  \Big) $, and define a new spinor $ \tilde{\Psi}_d = \tilde{U}_d \Psi_d $. The transformed equation becomes
\begin{equation}\label{dirac_eigen_eq_1}
i\partial_t \tilde{\Psi}_d=\tilde{H}_d \tilde{\Psi}_d .
\end{equation}
One can simplify the equation and can now be expressed in matrix form as
\begin{equation}
\begin{bmatrix}
i\partial_t  & i\sigma^3 \left(\frac{1}{4}\frac{f^{\prime}}{f}+\partial_{r^{\ast}}+\frac{f}{r}\right) +i\frac{f^{\frac{1}{2}}}{r} \hat{S} \\
i\sigma^3 \left(\frac{1}{4}\frac{f^{\prime}}{f}+\partial_{r^{\ast}}+\frac{f}{r}\right) +i\frac{f^{\frac{1}{2}}}{r} \hat{S}& i\partial_t 
\end{bmatrix} \tilde{\Psi}_d=0,
\end{equation}
where the operator $ \hat{S} $ is defined as
\begin{equation}
\hat{S} = \sigma^1 \left( \partial_\theta + \frac{\cot\theta}{2} \right) + \frac{\sigma^2}{\sin\theta} \partial_\phi.
\end{equation}
To further simplify the analysis, we assume a separable ansatz for the spinor field $\tilde{\Psi}$ as
\begin{equation}
\tilde{\Psi}_d=
\begin{bmatrix}
R_1(t,r^{\ast}) \chi(\theta,\phi) \\
-i\sigma^3 R_2(t,r^{\ast}) \chi(\theta,\phi)
\end{bmatrix},
\end{equation}
where, $\chi(\theta,\phi)$ is a column vector of two components. Substituting this form into the matrix equation yields a pair of coupled equations
\begin{equation}
\begin{aligned}
i\partial_t R_1 \chi + \left( \frac{1}{4} \frac{f'}{f} + \partial_{r^*} + \frac{f}{r} \right) R_2 \chi + \frac{f^{1/2}}{r} R_2 \hat{S} \sigma^3 \chi &= 0, \\
i\partial_t R_2 \chi - \left( \frac{1}{4} \frac{f'}{f} + \partial_{r^*} + \frac{f}{r} \right) R_1 \chi + \frac{f^{1/2}}{r} R_1 \hat{S} \sigma^3 \chi &= 0,
\end{aligned}
\end{equation}
where we have used $\sigma^3 \hat{S} = -\hat{S} \sigma^3$. To isolate the angular dependence, we define the eigenvalue equation $\hat{S} \sigma^3 \chi = \lambda \chi$, where $ \lambda = \pm (j + \frac{1}{2}) $ with $ l = j \pm \frac{1}{2} $. Here, $ j $ and $ l $ are the total and orbital angular momenta respectively. This leads to the final form of the two radial equations
\begin{equation}\label{Radial_R}
\begin{aligned}
& i\partial_t R_1 +\partial_{r^{\ast}} R_2 +\left(\frac{1}{4}\frac{f^{\prime}}{f}+\frac{f^{\frac{1}{2}}}{r}\lambda+\frac{f}{r}\right)R_2  =0,  \\
& i\partial_t R_2 -\partial_{r^{\ast}} R_1 -\left(\frac{1}{4}\frac{f^{\prime}}{f}-\frac{f^{\frac{1}{2}}}{r}\lambda+\frac{f}{r}\right)R_1 =0  .
\end{aligned}
\end{equation}
Now we proceed to solve the angular equation $\hat{S}\sigma^3\chi=\lambda\chi$ by taking the following ansatz for the angular spinor
\begin{equation}
\chi(\theta,\phi)=\frac{1}{\sqrt{2\pi}}e^{im\phi}
\begin{bmatrix}
\chi^+(\theta) \\
\chi^-(\theta)
\end{bmatrix},
\end{equation}
where $m \in | \mathbb{Z} | + \frac{1}{2} $ is the half-integer azimuthal quantum number. Substituting this into the angular equation yields two second-order differential equations for the functions $ \chi^{\pm}(\theta) $ as
\begin{equation}
\left[\frac{d^2}{d\theta^2}+\cot\theta\frac{d}{d\theta}-\frac{1}{\sin^2\theta}\left(m^2 \mp m\cos\theta+\frac{1}{4}  \right)-\frac{1}{4}+\lambda \right]\chi^{\pm}(\theta)=0 .
\end{equation}
These equations can be solved in terms of Jacobi polynomials. The general solution is given by
\begin{equation}
\chi^{\pm}(\theta) = A_n \, (1 - \cos\theta)^{\frac{1}{2}(m \mp \frac{1}{2})} (1 + \cos\theta)^{\frac{1}{2}(m \pm \frac{1}{2})} P^{(m \mp \frac{1}{2}, \, m \pm \frac{1}{2})}_{j - m}(\cos\theta),
\end{equation}
where $A_n = 2^{-\frac{1}{2}- m} \sqrt{\frac{(j - m)! (j + m)!}{(j - 1/2)!}}$ is a normalization constant and $ P^{(a,b)}_n(x) $ denotes the Jacobi polynomial of order $ n $ with parameters $a $ and $ b $. Consequently, the full angular spinor takes the form
\begin{equation}
\chi(\theta,\phi) = \frac{A_n}{\sqrt{2\pi}} e^{im\phi}
\begin{bmatrix}
(1 - \cos\theta)^{\frac{1}{2}(m - \frac{1}{2})} (1 + \cos\theta)^{\frac{1}{2}(m + \frac{1}{2})} P^{(m - \frac{1}{2}, \, m + \frac{1}{2})}_{j - m}(\cos\theta) \\
(1 - \cos\theta)^{\frac{1}{2}(m + \frac{1}{2})} (1 + \cos\theta)^{\frac{1}{2}(m - \frac{1}{2})} P^{(m + \frac{1}{2}, \, m - \frac{1}{2})}_{j - m}(\cos\theta)
\end{bmatrix}.
\end{equation}
Now we can write the four component spinor in Cartesian gauge as $\Psi_c = S \Psi_d = e^{-\frac{1}{2}\phi \gamma_d^1 \gamma_d^2} \\e^{-\frac{1}{2}\theta \gamma_d^3 \gamma_d^1} \tilde{\Psi}_d $. Therefore
\begin{equation}
\Psi_c = \begin{bmatrix}
R_1 \frac{1}{\sqrt{2\pi}}e^{im\phi}e^{-\frac{i}{2}\phi\sigma^3}e^{-\frac{i}{2}\theta\sigma^2}\begin{pmatrix}
\chi^+\\
\chi^-
\end{pmatrix} \\
-i R_2 \frac{1}{\sqrt{2\pi}}e^{im\phi}e^{-\frac{i}{2}\phi\sigma^3}e^{-\frac{i}{2}\theta\sigma^2}\begin{pmatrix}
\chi^+\\
-\chi^-
\end{pmatrix}
\end{bmatrix} .
\end{equation}
Using the properties of the associated Legendre and the Jocobi polynomials, one can show that~\cite{thaller2005advanced}
\begin{equation}
\begin{aligned}
\frac{1}{\sqrt{2\pi}}e^{im\phi}e^{-\frac{i}{2}\phi\sigma^3}e^{-\frac{i}{2}\theta\sigma^2}\begin{pmatrix}
\chi^+\\
\chi^-
\end{pmatrix}
&= \frac{(-2)^{m+1/2} (j - 1/2)!}{\sqrt{(j + m)! (j -m)!}} \, \mathcal{Y}^{j m}_{j+1/2}(\theta,\phi) , \\
\frac{1}{\sqrt{2\pi}}e^{im\phi}e^{-\frac{i}{2}\phi\sigma^3}e^{-\frac{i}{2}\theta\sigma^2}\begin{pmatrix}
\chi^+\\
- \chi^-
\end{pmatrix}
&= - \frac{(-2)^{m+1/2} (j - 1/2)!}{\sqrt{(j + m)! (j - m)!}} \, \mathcal{Y}^{j m}_{j-1/2}(\theta,\phi) ,
\end{aligned}
\end{equation}
where $\mathcal{Y}^{j\,m}_{j\pm\frac{1}{2}}$ are the spinor spherical harmonics given by
\begin{equation}
\begin{aligned}
\mathcal{Y}^{j\,m}_{j+\frac{1}{2}} &=\frac{1}{\sqrt{2j+2}}
\begin{pmatrix}
-\sqrt{j-m+1}\,Y_{j+\frac{1}{2},m-\frac{1}{2}} \\
\sqrt{j+m+1}\,Y_{j+\frac{1}{2},m+\frac{1}{2}}
\end{pmatrix} , \\
\mathcal{Y}^{j\,m}_{j-\frac{1}{2}} &=\frac{1}{\sqrt{2j}}
\begin{pmatrix}
\sqrt{j+m}\,Y_{j-\frac{1}{2},m-\frac{1}{2}} \\
\sqrt{j-m}\,Y_{j-\frac{1}{2},m+\frac{1}{2}}
\end{pmatrix} ,
\end{aligned}
\end{equation}
and $Y_{j \pm 1/2, m \pm 1/2}(\theta,\phi)$ are the spherical harmonics. To analyze the radial equations~\eqref{Radial_R}, we transform them into a pair of Schrödinger-like wave equations. This is achieved by employing the ansatz
\begin{equation}
    R_1(t,r^*) = e^{-i\omega t} \frac{U_1(r^*)}{rf^{1/4}}, \quad R_2(t,r^*) = e^{-i\omega t} \frac{U_2(r^*)}{rf^{1/4}}.
\end{equation}
This substitution decouples the original system and yields two independent wave equations for the new functions $U_1(r^*)$ and $U_2(r^*)$ as
\begin{equation}
\begin{aligned}
    \frac{d^2 U_1}{d{r^*}^2}+\left( \omega^2 - V^{(1)}_{\rm eff}\right)U_1 &= 0, \\
    \frac{d^2 U_2}{d{r^*}^2}+ \left( \omega^2 - V^{(2)}_{\rm eff}\right)U_2 &= 0,
\end{aligned}
\end{equation}
where $\omega$ is the mode frequency. The effective potentials $V^{(1,2)}_{\rm eff}$ are functions of the radial coordinate $r$ and are given by:
\begin{equation}
\begin{aligned}
    V^{(1)}_{\rm eff}(r) &= -\frac{\lambda(\sqrt{f}-\lambda)}{r^2} - \frac{r_g \lambda}{2r^3}(2\lambda - 3\sqrt{f}), \\
    V^{(2)}_{\rm eff}(r) &= \frac{\lambda(\sqrt{f}+\lambda)}{r^2} - \frac{r_g \lambda}{2r^3}(2\lambda + 3\sqrt{f}).
\end{aligned}
\end{equation}
In the asymptotic limits, as $r \to \infty$ ($r^* \to \infty$) and $r \to r_g$ ($r^* \to -\infty$), both effective potentials vanish. Consequently, the radial solutions $U_{1,2}$ reduce to simple plane waves of the form $e^{\pm i \omega r^*}$. Reconstructing the full spinor from these radial parts and the corresponding angular spherical harmonics, the asymptotic behavior of fermionic fields in the cartesian gauge is found to be
\begin{equation}
    \Psi_c \sim \frac{1}{r} e^{- i\omega t} e^{\pm i \omega r^{\ast}}
    \begin{bmatrix}
     \mathcal{Y}^{j\,m}_{j+\frac{1}{2}}(\theta,\phi) \\
    i \mathcal{Y}^{j\,m}_{j-\frac{1}{2}}(\theta,\phi)
    \end{bmatrix}.
\end{equation}
Similarly, to the scalar field, we now proceed to quantize the scalar field in the asymptotic limit. On past null infinity, $\mathscr{I}^-$, the field operator can be decomposed in terms of a basis of ingoing positive- and negative-frequency modes
\begin{equation}
    \hat{\Psi} = \int_0^{\infty} d\omega \sum_{l,m} \left( f_{\omega lm}\hat{a}^-_{\omega lm} + g_{\omega lm}\hat{b}^{+}_{\omega lm} \right)
\end{equation}
The mode functions $\{f_{\omega lm}\}$ and $\{g_{\omega lm}\}$ represent incoming particles and anti-particles respectively. They are defined by
\begin{align}
    f_{\omega lm} = \frac{1}{\sqrt{2\pi}}\frac{1}{r}e^{-i\omega v}\varphi^+_{lm}(\theta,\phi) , \qquad
    g_{\omega lm} = \frac{1}{\sqrt{2\pi}}\frac{1}{r}e^{+i\omega v}\varphi^-_{lm}(\theta,\phi)
\end{align}
where $v=t+r^*$ is the advanced null coordinate. The angular dependence is contained within the spinor harmonics $\varphi^\pm_{lm}$ as
\begin{align}
    \varphi^+_{lm} = \begin{bmatrix}
     \mathcal{Y}^{j\,m}_{j+\frac{1}{2}} \\
    0 
    \end{bmatrix}, \qquad
    \varphi^-_{lm} = \begin{bmatrix}
     0 \\
    \mathcal{Y}^{j\,m}_{j-\frac{1}{2}} 
    \end{bmatrix}.
\end{align}
The operators $\hat{a}^-_{\omega lm}$ and $\hat{b}^-_{\omega lm}$ are the annihilation operators for particles and anti-particles, respectively. Together with their corresponding creation operators, $\hat{a}^{+}_{\omega lm}$ and $\hat{b}^{+}_{\omega lm}$, they satisfy the standard fermionic anti-commutation relations
\begin{align}\label{sch_fermion_ACR_negative}
    \left\{ \hat{a}^-_{\omega lm}, \hat{a}^{+}_{\omega^{\prime} l^{\prime}m^{\prime}} \right\} = \delta(\omega-\omega^{\prime})\delta_{ll^{\prime}}\delta_{m m^{\prime}}, \qquad
    \left\{ \hat{b}^-_{\omega lm}, \hat{b}^{+}_{\omega^{\prime} l^{\prime}m^{\prime}} \right\} = \delta(\omega-\omega^{\prime})\delta_{ll^{\prime}}\delta_{m m^{\prime}}.
\end{align}
All other anti-commutators vanish. The ingoing vacuum state, $\vert 0_-\rangle$, on $\mathscr{I}^-$ is defined as the state annihilated by all ingoing particle and anti-particle operators
\begin{equation}
    \hat{a}^-_{\omega lm} \vert 0_-\rangle = 0 \quad \text{and} \quad \hat{b}^-_{\omega lm} \vert 0_-\rangle = 0 \quad \forall \, \omega, l, m.
\end{equation}
The mode functions form a complete orthonormal set on $\mathscr{I}^-$ with respect to the conserved inner product, satisfying the relations
\begin{equation}
\begin{aligned}
    \left(f_{\omega lm},f_{\omega^{\prime} l^{\prime}m^{\prime}} \right)_{\mathscr{I}^-} &= \delta(\omega-\omega^{\prime})\delta_{ll^{\prime}}\delta_{m m^{\prime}} , \quad
    \left(g_{\omega lm},g_{\omega^{\prime}l^{\prime}m^{\prime}} \right)_{\mathscr{I}^-} = \delta(\omega-\omega^{\prime})\delta_{ll^{\prime}}\delta_{m m^{\prime}}, \\
    \left(f_{\omega lm}, g_{\omega^{\prime}l^{\prime}m^{\prime}} \right)_{\mathscr{I}^-} &= 0.
\end{aligned}
\end{equation}
The spinorial inner product defined for the modes belonging to the space of solutions of the Dirac equation on a Cauchy surface $\Sigma$ is given by:
\begin{equation}
    (\Psi_1,\Psi_2) = \int_{\Sigma} \sqrt{-g} \,d^3 x\, \bar{\Psi}_1 \gamma^\mu n_\mu \Psi_2 
\end{equation}
where $n_\mu$ is the future-directed unit normal vector to the hypersurface. The Dirac adjoint in this formalism is defined as $\bar{\Psi} = \Psi^\dagger \alpha$, with $\alpha = -\gamma^0$.\\

Similarly, the fermionic field operator on future null infinity, $\mathscr{I}^+$, can be decomposed. A complete basis at $\mathscr{I}^+$ must include both outgoing modes, which propagate to future infinity, and ingoing modes, which are lost to the BH event horizon. The field operator is therefore written in terms of four sets of mode coefficients
\begin{equation}
    \hat{\Psi}=\int_0^{\infty} d\omega \sum_{l,m}\left( p_{\omega lm}\hat{c}^-_{\omega lm}+q_{\omega lm}\hat{d}^{+}_{\omega lm}+r_{\omega lm}\hat{h}^-_{\omega lm}+s_{\omega lm}\hat{k}^{+}_{\omega lm} \right).
\end{equation}
Here, $\{p_{\omega lm}\}$ and $\{q_{\omega lm}\}$ are the outgoing mode functions, while $\{r_{\omega lm}\}$ and $\{s_{\omega lm}\}$ represent the ingoing modes that are absorbed by the BH. The outgoing modes are defined in terms of the retarded null coordinate, $u=t-r^*$ as
\begin{align}
    p_{\omega lm} =\frac{1}{\sqrt{2\pi}}\frac{1}{r}e^{-i\omega u}\varphi^+_{lm} , \qquad
    q_{\omega lm} =\frac{1}{\sqrt{2\pi}}\frac{1}{r}e^{+i\omega u}\varphi^-_{lm}.
\end{align}
The field is quantized by imposing canonical anti-commutation relations. The outgoing particle operators ($\hat{c}$) and anti-particle operators ($\hat{d}$) satisfy
\begin{align}\label{sch_fermion_ACR_positive}
    \left\{ \hat{c}^-_{\omega lm},\hat{c}^{+}_{\omega^{\prime} l^{\prime}m^{\prime}} \right\} =\delta(\omega-\omega^{\prime})\delta_{ll^{\prime}}\delta_{m m^{\prime}} , \quad
    \left\{ \hat{d}^-_{\omega lm},\hat{d}^{+}_{\omega^{\prime} l^{\prime}m^{\prime}} \right\} =\delta(\omega-\omega^{\prime})\delta_{ll^{\prime}}\delta_{m m^{\prime}}.
\end{align}
The ingoing particle ($\hat{h}$) and anti-particle ($\hat{k}$) operators also satisfy the identical set of anti-commutation relations. All other anti-commutators vanish. Corresponding to these operators, it is useful to define the vacuum state on future null infinity, $\vert 0_+\rangle$, often called the Unruh vacuum, which is defined as the state that is annihilated by all ingoing and outgoing annihilation operators
\begin{equation}
    \hat{c}^-_{\omega lm} \vert 0_+\rangle = \hat{d}^-_{\omega lm} \vert 0_+\rangle = \hat{h}^-_{\omega lm} \vert 0_+\rangle = \hat{k}^-_{\omega lm} \vert 0_+\rangle = 0 \quad \forall \, \omega, l, m.
\end{equation}

\subsubsection*{Particle Creation}
To calculate the particle spectrum for fermions, we follow the same procedure to that of the scalar field case. The purely positive-frequency outgoing mode, $p_{\omega lm}$, at future null infinity ($\mathscr{I}^+$) is selected to propagated backward in time along null geodesics $\gamma$, as depicted in Fig.~\ref{Penrose_diagram}. Due to the scattering off the spacetime curvature, the corresponding mode on past null infinity ($\mathscr{I}^-$), denoted $p^{(2)}_{\omega lm}$, is found to be a superposition of positive and negative frequencies. Its form is given by
\begin{equation}
    p^{(2)}_{\omega lm} \approx \frac{1}{\sqrt{2\pi}} \frac{1}{r}  e^{-i\omega u(v)} \varphi^+_{lm},
\end{equation}
where the relationship between the retarded time $u$ and the advanced time $v$ is given by \eqref{u_v_rel_sch}. This mode mixing is formalized by expressing the scattered mode $p^{(2)}_{\omega lm}$ as a Bogoliubov transformation of the basis modes on $\mathscr{I}^-$. It is therefore expanded as
\begin{equation}
    p_{\omega l m}^{(2)} =\int_0^{\infty}d\omega^{\prime} \sum_{l^{\prime}, m^{\prime}} \left(\alpha_{\omega l m\, \omega^{\prime} l^{\prime} m^{\prime}}f_{\omega^{\prime}l^{\prime} m^{\prime}}+\beta_{\omega lm\, \omega^{\prime}l^{\prime} m^{\prime}}f^{*}_{\omega^{\prime}l^{\prime} m^{\prime}} \right),
\end{equation}
where the Bogoliubov coefficients, $\alpha_{\omega l m\, \omega^{\prime} l^{\prime} m^{\prime}}$ and $\beta_{\omega l m\, \omega^{\prime} l^{\prime} m^{\prime}}$, are determined by taking the inner product on $\mathscr{I}^-$ as
\begin{equation}\label{bogo_sch_fermion}
\begin{aligned}
\alpha_{\omega lm \; \omega^{\prime}l^{\prime} m^{\prime}} &=\left(p_{\omega lm}^{(2)},f_{\omega^{\prime}l^{\prime} m^{\prime}} \right)_{\mathscr{I}^-} \approx \frac{e^{-i\omega^\prime v_0}}{2\pi}  K^{i\frac{\omega}{\kappa}}  \left( -i \omega^\prime \right)^{\left(-1+i\frac{\omega}{\kappa}\right)}   \Gamma\left(1-i\frac{\omega}{\kappa}\right)  \delta_{ll^{\prime}}\delta_{mm^{\prime}},  \\
\beta_{\omega lm \; \omega^{\prime}l^{\prime} m^{\prime}} &=-\left(p_{\omega lm}^{(2)},f^{\ast}_{\omega^{\prime}l^{\prime} m^{\prime}} \right)_{\mathscr{I}^-} \approx - \alpha_{\omega l m \; (-\omega^\prime) l^\prime m^\prime}. 
\end{aligned}
\end{equation}
The canonical anti-commutation relations for the field operators impose a normalization condition on the Bogoliubov coefficients
\begin{equation}
    \int_0^{\infty} d\omega^{\prime}\sum_{l^{\prime},m^{\prime}}\left( |\alpha_{\omega lm\, \omega^{\prime}l^{\prime}m^{\prime}}|^2 + |\beta_{\omega lm\, \omega^{\prime}l^{\prime}m^{\prime}}|^2 \right)= \Gamma_{\omega lm} \delta(0).
\end{equation}
The number of created particles in a given mode $(\omega, l, m)$, as measured by an observer at $\mathscr{I}^+$, is the expectation value of the outgoing number operator in the ingoing vacuum state, $| 0_- \rangle$. This is given by
\begin{equation}
    \langle 0_- | \hat{c}_{\omega lm}^{\dagger(2)} \hat{c}_{\omega lm}^{-(2)} | 0_- \rangle = \int_0^\infty d\omega' \sum_{l',m'} |\beta_{\omega lm\, \omega' l' m'}|^2 = \frac{\Gamma_{\omega lm}}{e^{\frac{2\pi\omega}{\kappa}}+1}\delta(0),
\end{equation}
where $\hat{c}_{\omega lm}^{-(2)} = (p_{\omega lm}^{(2)}, \hat{\Psi})$ and $\hat{c}_{\omega lm}^{+(2)} = (p_{\omega lm}^{(2)*}, \hat{\Psi})$ are the annihilation and creation operators associated with the scattered mode $p^{(2)}_{\omega lm}$. Therefore the number of particles emitted per unit time, per unit frequency, is a Fermi-Dirac distribution modified by the graybody factor
\begin{equation}
    n_{\omega} = \frac{\Gamma_{\omega}}{e^{\,\omega / T_{\rm BH}} + 1},
\end{equation}
where $\Gamma_\omega = \sum_{l,m} \Gamma_{\omega lm}$ is the total graybody factor for the fermionic field.

In the subsequent section, we apply the preceding formalism to calculate the Hawking radiation spectrum for the Kerr geometry. Our analysis will encompass the emission of both scalar and fermionic quantum fields to provide a comprehensive treatment with details not usually available in the literature.


\subsection{Kerr black holes}

Next, we consider the case of a rotating BH, described by the Kerr solution to Einstein's field equations. The Kerr BH represents a more realistic astrophysical BH, as most BHs are expected to possess angular momentum due to their formation history. The spacetime geometry surrounding such a rotating body of mass $M$ and angular momentum $J$ is given by the Kerr metric. In Boyer--Lindquist coordinates, the metric reads~\cite{PhysRevLett.11.237, Boyer:1966qh}
\begin{align}\label{kerr_metric}
ds^2 =& -\left(1-\frac{r r_g}{\Sigma^2}  \right)dt^2+\frac{\Sigma^2}{\Delta}dr^2+\Sigma^2 d\theta^2+\left\lbrace (r^2+a_k^2)\sin^2\theta+\frac{a_k^2 r r_g \sin^4 \theta}{\Sigma^2} \right\rbrace d\phi^2 \notag \\
&-\frac{2a_k \, r \, r_g \, \sin^2 \theta}{\Sigma^2}d\phi dt
\end{align}
where $r_g=2GM$ is the Schwarzschild radius, $\Sigma^2 = r^2 + a_k^2 \cos^2 \theta$, and $\Delta = r^2 - r_g r + a_k^2$. The metric determinant is given by $\sqrt{-g} = \Sigma^2 \sin\theta$. We define the specific angular momentum of the BH as $a_k = J/M$, and the dimensionless spin parameter is $a_* = a_k / GM$. The event horizons are located at
\begin{equation}
r_\pm = GM\left(1 \pm \sqrt{1 - a_*^2}\right).
\end{equation}
The Kerr geometry introduces frame dragging, where the spacetime itself is dragged in the direction of the BH's spin. This effect is encoded in the off-diagonal term $d\phi dt$ in the metric. Such a spacetime alters the dynamics of fields near the BH and modifies the structure of quantum radiation.
\subsubsection{Scalar fields}
To analyze quantum field behavior in this rotating background, we consider a massless scalar field $\Phi$ governed by the Klein--Gordon equation, $\Box \Phi = 0$. Assuming a separable ansatz for the scalar field,
\begin{equation}
\Phi(t,r,\theta,\phi) =\mathcal{R}(r) S(\theta) e^{i m \phi} e^{-i \omega t},
\end{equation}
the Klein-Gordon equation separates into radial and angular parts~\cite{PhysRevD.12.2963}
\begin{align}\label{R_eq}
\Delta \partial_{r}\left( \Delta \partial_{r}\mathcal{R}\right)+\big\lbrace  \omega^2(r^2+a_k^2)^2 -2m\omega a_k r r_g +m^2 a_k^2  -\Delta(a_k^2\omega^2+\lambda_l) \big\rbrace \mathcal{R}=0
\end{align}
and
\begin{align}\label{S_eq}
\frac{1}{\sin\theta}\partial_{\theta}\left( \sin \theta \partial_{\theta}S\right)+ \left\lbrace a_k^2\omega^2\cos^2\theta -\frac{m^2}{\sin^2\theta}+\lambda_l \right\rbrace S=0.
\end{align}
The solution of Eq.~(\ref{S_eq}) is the oblate spherodial harmonics~\cite{2003JPhA...36.5477F} $S_{lm}(ia_k \omega,\cos\theta)$ with eigen value $\lambda_l$ where $l,m$ are integers with $ \vert m \vert \leq l $. In the limit $a_k \rightarrow 0 $, the $S_{lm}$ reduces to $P_{lm}(\cos \theta)$, associated Legendre functions and $\lambda_l$ becomes $l(l+1)$. By redefining the radial function via a tortoise coordinate \( r^* \) and \( U(r) = \mathcal{R}(r) \sqrt{r^2 + a_k^2} \), Eq.~(\ref{R_eq}) can be recast as a Schr\"odinger like equation
\begin{equation}
\frac{d^2U}{d{r^{\ast}}^2}+V_{\rm eff}(r)U=0
\end{equation}
where \( dr^* / dr =(r^2 + a_k^2)/\Delta  \) and the effective potential is
\begin{align}
V_{\rm eff}(r)=& \omega^2+(r^2+a_k^2)^{-2}\big\lbrace m^2a_k^2-2m \omega a_k r r_g -\Delta\big( l(l+1) +\omega^2a_k^2 \big) \big\rbrace \notag \\
& -\Delta(r^2+a_k^2)^{-3}(\Delta+r(2r-r_g)) +3r^2\Delta^2(r^2+a_k^2)^{-4}.
\end{align}
In the asymptotic limit \( r \rightarrow \infty \) (\( r^* \rightarrow \infty \)), the potential approaches \( V_{\rm eff} \rightarrow \omega^2 \), yielding
\begin{equation}
\mathcal{R}(r) \sim \frac{1}{r} e^{\pm i \omega r^*}.
\end{equation}
Near the event horizon $r \rightarrow r_+ $ ($ r^* \rightarrow -\infty $), the potential approaches $ V_{\rm eff} \rightarrow -(\omega - m \Omega_h)^2 $, where $ \Omega_h = a_k/r_g r_+ $, leading to
\begin{equation}
\mathcal{R}(r) \sim \frac{1}{r} e^{\pm i (\omega - m \Omega_h) r^*}.
\end{equation}
Thus, the asymptotic behavior of the scalar field solution is
\begin{equation}
\Phi(t,r,\theta,\phi) \sim 
\begin{cases}
\frac{1}{r}e^{\pm i \omega r^*}e^{-i \omega t} e^{i m \phi} S_{lm}, & r^* \rightarrow \infty, \\
\frac{1}{r}e^{\pm i \tilde{\omega} r^*}e^{-i \omega t} e^{i m \phi} S_{lm}, & r^* \rightarrow -\infty,
\end{cases}
\end{equation}
where $ \tilde{\omega} = \omega - m \Omega_h $. The frame-dragging effects of the Kerr background result in a shift of the effective frequency observed by an asymptotic observer. Specifically, modes of the form $e^{-i\omega t + im\phi}$ experience a frequency shift to $\omega - m\Omega_h$. This shift is central to understanding superradiance from Kerr BHs. In the next part, we use these mode functions to quantize the field and derive the standard Hawking radiation spectrum for the Kerr BH.

Similar to that of the Schwarzschild black hole, in the past null infinity $\mathscr{I}^-$, the field can be decomposed as
\begin{equation}
\hat{\Phi}=\int_0^{\infty} d\omega \sum_{l,m} \left(f_{\omega l m}\hat{a}^-_{\omega l m}+f^{\ast}_{\omega l m}\hat{a}^+_{\omega l m} \right)
\end{equation}
where, the form of $\lbrace f_{\omega l m}\rbrace $ are given by
\begin{equation}\label{f_form}
f_{\omega l m}=\frac{1}{\sqrt{2\pi \omega}}\frac{1}{r}e^{-i\omega v}e^{i m \phi} S_{lm}(ia_k \omega ,\cos\theta),
\end{equation}
The modes $\lbrace f_{\omega l m}\rbrace $ forms a complete orthonormal set on $\mathscr{I}^-$ as that in Schwarzschild case given by Eq.~(\ref{normal_eq}). The quantum field on $\mathscr{I}^+$ can be written interms of ingoing and outgoing modes as
\begin{align}
\hat{\Phi}=\int_0^{\infty} d\omega \sum_{l, m} \big(  p_{\omega l m}\hat{b}^-_{\omega l m}+p^{\ast}_{\omega l m} \hat{b}^+_{\omega l m} +q_{\omega l m}\hat{c}^-_{\omega l m}  +q^{\ast}_{\omega l m}\hat{c}^+_{\omega l m} \big)
\end{align}
where, $\lbrace p_{\omega l m}\rbrace$ and $\lbrace q_{\omega l m}\rbrace$ are the outgoing and ingoing modes respectively. The form of $\lbrace p_{\omega l m}\rbrace$ on $\mathscr{I}^+$ is given by
\begin{equation}\label{p_form}
p_{\omega l m}=\frac{1}{\sqrt{2\pi \omega}}\frac{1}{r}e^{-i\omega u}e^{i m \phi} S_{lm}(ia_k \omega ,\cos\theta).
\end{equation}
Here $\hat{b}^-_{\omega l m}$ and $\hat{b}^+_{\omega l m}$ are the annihilation and creation operators respectively for outgoing modes. Those modes $\lbrace p_{\omega l m}\rbrace$ and $\lbrace q_{\omega l m}\rbrace$ satisfies the orthonormality condition on future null infinity $\mathscr{I}^+$ and past event horizon $\mathcal{H}^-$ respectively. The vacuum $\vert 0_+\rangle $ on $\mathscr{I}^+$ is defined as $\hat{b}^-_{\omega l m} \vert 0_+\rangle=0$ or $\hat{c}^-_{\omega l m} \vert 0_+\rangle=0$.\\

\subsubsection*{Particle Creation for scalar modes }
Similarly, proceeding as earlier we now trace the mode $ p_{\omega l m} $ from future null infinity $ \mathscr{I}^+ $ back to past null infinity $ \mathscr{I}^- $ along the null path $ \gamma $ and during this process, $ p_{\omega l m}^{(1)} $ fraction of the mode enters the BH and $ p_{\omega l m}^{(2)} $, is scattered and reaches $ \mathscr{I}^- $. The form of $ p_{\omega l m}^{(2)} $ at $ \mathscr{I}^- $ is then expressed as
\begin{equation}
p_{\omega l m}^{(2)} \approx \frac{1}{\sqrt{2\pi \omega}}\frac{1}{r}e^{-i\tilde{\omega} u(v) }e^{i m \tilde{\phi}_+} S_{lm}(ia_k \omega ,\cos\theta)
\end{equation}
where $\tilde{\phi}_+=\phi-\Omega_h t_0 $ is the azimuthal angular coordinate far outside the collapsing body at some early time $t_0$ and is continous at the horizon $r_+$~\cite{Parker:2009uva}. For a Kerr BH, the coordinate transformation $ u(v) $ is approximated by
\begin{equation}\label{u_v_relation_kerr}
u(v) \approx - \frac{1}{\kappa} \ln \left[\frac{v_0-v}{K}\right],
\end{equation}
where $\kappa = (r_+ - r_-)/2(r_+^2 + a_k^2)$ is the surface gravity of the Kerr BH. Now the Bogoliubov coefficients can be calculated as
\begin{equation}\label{bogo_kerr_boson}
\begin{aligned}
\alpha_{\omega lm \; \omega^{\prime}l^{\prime} m^{\prime}} & \approx \frac{1}{4\pi} \sqrt{\frac{\omega^\prime}{\omega}} e^{-i(m\Omega_h t_0-\omega^{\prime}v_0) } K^{-i\frac{\tilde{\omega}}{\kappa}}    (i\omega^{\prime})^{-\left(1+i\frac{\tilde{\omega}}{\kappa}\right)}\Gamma\left(1+i\frac{\tilde{\omega}}{\kappa}\right) \delta_{l l^{\prime}} \delta_{m m^{\prime}} ,\\
\beta_{\omega lm \; \omega^{\prime}l^{\prime} m^{\prime}} & \approx -i \alpha_{\omega l m \; (-\omega^{\prime}) l^{\prime} m^{\prime}}.
\end{aligned}
\end{equation}
From here on can derive the relation between $\alpha_{\omega lm \; \omega^{\prime}l^{\prime} m^{\prime}}$ and $\beta_{\omega lm \; \omega^{\prime}l^{\prime} m^{\prime}}$ as
\bea
\vert \beta_{\omega l m \; \omega^{\prime} l^{\prime} m^{\prime}}\vert ^2=e^{-\frac{2\pi \tilde{\omega} }{\kappa}}\vert \alpha_{\omega l m \; \omega^{\prime} l^{\prime} m^{\prime}}\vert ^2 . \label{alpha_beta_relation_kerr}
\eea
Using Eq.~(\ref{normal_alpha_beta}), the total number of particles created in the mode $(\omega, l, m)$ will be
\begin{equation}\label{kerr_spectrum}
\langle_-0\vert \hat{b}_{\omega lm}^{+(2)}\hat{b}_{\omega lm}^{-(2)}\vert 0_-\rangle = \frac{\Gamma_{\omega lm}}{e^{\frac{2\pi \tilde{\omega} }{\kappa}}-1}\delta(0),
\end{equation}
or in a mode of frequency $\omega$, number density of incoming particle is
\begin{equation}
n_{\omega} = \sum_{l,m} \frac{\Gamma_{\omega lm}}{e^{\frac{ \omega-m\Omega_h }{T_{\rm BH}}}-1}.
\end{equation}
This is the expected Hawking radiation spectrum for a Kerr BH, with the characteristic frequency shift $\omega-m\Omega_h$, in contrast to the Schwarzschild case.

\subsubsection{Fermionic fields} 
We now turn to the analysis of Dirac fermions in the Kerr geometry. The evolution of the fermionic field, $\Psi$, is governed by the covariant Dirac equation $i\gamma^\mu \nabla_\mu \Psi = 0,$ where $\nabla_\mu$ is the spinor covariant derivative, as discussed in Section~\ref{sch_fermion_section}. Utilizing the tetrads and spin connection for the Kerr metric, the explicit form of this equation is given by
\begin{align}
    \Biggl\{ & \left[ \frac{r^2+a_k^2}{\sqrt{\Delta}}\gamma^0 + a_k\sin\theta\gamma^2 \right]\partial_t 
    + \sqrt{\Delta}\gamma^3\partial_r 
    + \gamma^1\partial_\theta 
    + \left[ \frac{a_k}{\sqrt{\Delta}}\gamma^0 + \frac{1}{\sin\theta}\gamma^2 \right]\partial_\phi \notag \\
    + & \frac{\cot\theta}{2\Sigma^2}(\Sigma^2-a_k^2\sin^2\theta)\gamma^1 +  \left[ \frac{r\sqrt{\Delta}}{\Sigma^2} + \frac{1}{4\Sigma^2\sqrt{\Delta}}\left\{r_g(r^2-a_k^2\cos^2\theta)-2r a_k^2\sin^2\theta\right\} \right]\gamma^3 \notag \\
    + & \frac{r_g a_k \sin\theta}{2\Sigma^2}i\gamma^1\gamma^5 
    - \frac{\sqrt{\Delta}a_k\cos\theta}{2\Sigma^2}i\gamma^5\gamma^3 
    \Biggr\}\Psi = 0.
\end{align} 
The matrix $\gamma^{5}$ denotes the chirality operator, defined as
\begin{equation}
\gamma^5 = \frac{i}{4!}\epsilon_{\mu \nu \sigma \delta} \gamma^\mu \gamma^\nu \gamma^\sigma \gamma^\delta ,
\end{equation}
where $\epsilon_{\mu\nu\sigma\delta}$ is the totally antisymmetric Levi-Civita tensor. The specific choice of gamma matrices and tetrads is detailed in Appendix~\ref{appendix_dirac_formalism}. To solve this partial differential equation, we employ a separation of variables. Following the standard procedure, we adopt a separable ansatz of the form~\cite{Chandrasekhar:1976ap, Unruh:1973bda, Page:1976jj, Teukolsky:1973ha}
\begin{equation}
    \Psi^L_\Lambda(t,r,\theta,\phi) = \frac{1}{\sqrt{8\pi^2} \mathcal{F}_L(r,\theta)} e^{-i\omega t} e^{im\phi} \begin{bmatrix} \eta^L_\Lambda(r,\theta) \\ L \eta^L_\Lambda(r,\theta) \end{bmatrix}.
\end{equation}
Here, $\Lambda \equiv \{\omega,l,m\}$ denotes the set of quantum numbers specifying the mode’s frequency and angular momenta. The label $L = +1$ corresponds to left-handed spinor modes, while $L = -1$ corresponds to right-handed ones, defined through $\gamma^5 \Psi^L_{\Lambda} = L \Psi^L_{\Lambda}$~\cite{Casals:2012es}. The function $\mathcal{F}_L(r,\theta)$ is given by
\begin{equation}
    \mathcal{F}_L(r,\theta)=\left[ \Delta(r - i a_k L \cos\theta)^2 \sin^2\theta \right]^{1/4}.
\end{equation}
The two-component spinor $\eta^L_\Lambda$ is itself assumed to be separable into radial and angular parts
\begin{equation}
    \eta^L_\Lambda = 
    \begin{bmatrix}
    {R_1}^L_{\Lambda}(r) {S_1}_\Lambda(\theta) \\
    {R_2}^L_{\Lambda}(r) {S_2}_\Lambda(\theta)
    \end{bmatrix}.
\end{equation}
Substituting this ansatz into the full Dirac equation allows the radial and angular variables to be separated, yielding two sets of coupled first-order ordinary differential equations. The radial equations are~\cite{Dolan:2015eua, Unruh:1974bw, Dai:2023zcj}
\begin{equation}
\begin{aligned}
    & \sqrt{\Delta}\left(\partial_r - \frac{iKL}{\Delta}\right){R_1}^L_{\Lambda}(r)=\lambda {R_2}^L_{\Lambda}(r), \\
    & \sqrt{\Delta}\left(\partial_r + \frac{iKL}{\Delta}\right) {R_2}^L_{\Lambda}(r)=\lambda {R_1}^L_{\Lambda}(r),
\end{aligned}
\end{equation}
where $K=(r^2+a_k^2)\omega -a_k m$, and $\lambda$ is the separation constant. In the Schwarzschild limit ($a_k \to 0$), this constant is related to the total angular momentum, $\lambda \to l+\frac{1}{2}$ with $l=\frac{1}{2},\frac{3}{2},\ldots$. The corresponding angular equations are given by
\begin{equation}
\begin{aligned}
    & \left[\partial_\theta + (a_k\omega\sin\theta - \frac{m}{\sin\theta})\right]{S_1}_{\Lambda}(\theta)= \lambda {S_2}_{\Lambda}(\theta), \\
    & \left[\partial_\theta - (a_k\omega\sin\theta - \frac{m}{\sin\theta})\right]{S_2}_{\Lambda}(\theta)=-\lambda {S_1}_{\Lambda}(\theta).
\end{aligned}
\end{equation}
The angular functions ${S_1}_{\Lambda}$ and ${S_2}_{\Lambda}$, which are solutions to the coupled angular equations and are real functions. They depend on the frequency $\omega$ and are chosen to be orthonormal according to the condition
\begin{equation}
\int_0^{\pi} {S_1}_{\{\omega l m\} } {S_1}_{\{\omega l^\prime m^\prime \} } d\theta=\int_0^{\pi} {S_2}_{\{ \omega l m\}} {S_2}_{\{\omega l^\prime m^\prime\}} d\theta= \delta_{l l^\prime} \delta_{m m^\prime} .
\end{equation}
The system of coupled first-order radial equations can be decoupled to yield two independent, second-order master differential equations for ${R_1}^L_{\Lambda}$ and ${R_2}^L_{\Lambda}$. These master equations can then be transformed into a standard Schr\"odinger-like form, which is more amenable to analysis. This is achieved by performing the following change of variables as
\begin{equation}
    {R_1}^L_{\Lambda}(r) = \frac{{U_1}^L_{\Lambda}(r^*)}{\sqrt{g(r)}}, \quad {R_2}^L_{\Lambda}(r) = \frac{{U_2}^L_{\Lambda}(r^*)}{\sqrt{g(r)}},
\end{equation}
where the function $g(r)$ used for generating the transformation is defined as $g(r)=(r^2+a_k^2)/ \sqrt{\Delta}$. This procedure leads to two decoupled wave equations for the functions ${U_1}^L_{\Lambda}(r^*)$ and ${U_2}^L_{\Lambda}(r^*)$ as
\begin{equation}
\begin{aligned}
    \frac{d^2{U_1}^L_{\Lambda}}{d{r^*}^2} + V_{\rm eff}^{(1L)}(r)  {U_1}^L_{\Lambda} &= 0, \\
    \frac{d^2{U_2}^L_{\Lambda}}{d{r^*}^2} + V_{\rm eff}^{(2L)}(r)  {U_2}^L_{\Lambda} &= 0,
\end{aligned}
\end{equation}
where the effective potentials, $V_{\rm eff}^{(1L)}$ and $V_{\rm eff}^{(2L)}$ are given by
\begin{equation}
\begin{aligned}
    V_{\rm eff}^{(1L)} &= \frac{1}{g^2} \left[ \frac{K^2}{\Delta}-\lambda^2+\frac{iKL}{2\Delta}(2r-r_g)-2iLr\omega  \right] -\frac{1}{2}\frac{d}{dr^*}\left( \frac{g'}{g} \right)-\frac{1}{4}\left( \frac{g'}{g} \right)^2, \\
    V_{\rm eff}^{(2L)}  &= \frac{1}{g^2} \left[\frac{K^2}{\Delta}-\lambda^2-\frac{iKL}{2\Delta}(2r-r_g)+2iLr\omega \right] -\frac{1}{2}\frac{d}{dr^*}\left( \frac{g'}{g} \right)-\frac{1}{4}\left( \frac{g'}{g} \right)^2.
\end{aligned}
\end{equation}
Here, prime denotes a derivative with respect to the tortoise coordinate, $g' \equiv dg/dr^*$. We now analyze the behavior of the radial solutions in the asymptotic limit of large $r$ (corresponding to $r^* \to +\infty$). In this region, the effective potentials approach a constant value, $V_{\rm eff}^{(1L,2L)} \to \omega^2$, the solutions are therefore plane waves, $U_{1,2} \sim e^{\pm i\omega r^*}$. This determines the asymptotic behavior of the two-component spinor $\eta^L_\Lambda$, which projects onto the appropriate angular function based on the helicity $L$ as~\cite{Vilenkin:1978is}
\begin{equation}
    \eta^L_{\Lambda} \sim e^{\pm i \omega r^*} \begin{bmatrix}
    (1+L) S_{1 \Lambda} (\theta) \\
    (1-L) S_{2 \Lambda} (\theta)
    \end{bmatrix}.
\end{equation}
In the same limit, the function $\mathcal{F}_L(r,\theta)$ from the ansatz simplifies to $\mathcal{F}_L(r,\theta) \to r\sqrt{\sin\theta}$. Combining these results allows us to construct the asymptotic form of the full four-component spinor solution at spatial infinity
\begin{equation}
    \Psi^L_{\Lambda}(t,r,\theta,\phi) \sim \frac{1}{r\sqrt{8 \pi^2 \sin\theta}} e^{-i \omega t} e^{\pm i \omega r^*} e^{im\phi} 
    \begin{bmatrix}
    (1+L) S_{1 \Lambda} (\theta) \\
    (1-L) S_{2 \Lambda} (\theta)  \\
    L(1+L) S_{1 \Lambda} (\theta) \\
    L(1-L) S_{2 \Lambda} (\theta)
    \end{bmatrix}.
\end{equation}
For the purposes of quantization, we require the decomposition of the field operator on past null infinity, $\mathscr{I}^-$. Following a procedure analogous to the Schwarzschild case, the field operator is expanded in terms of a complete set of ingoing modes
\begin{equation}
    \hat{\Psi}=\int_0^{\infty} d\omega \sum_{l=s}^{\infty} \sum_{m=-l}^{+l} \left( f_{\omega lm}\hat{a}^-_{\omega lm}+g_{\omega lm}\hat{b}^{\dagger}_{\omega lm} \right),
\end{equation}
where $s=1/2$ is the spin of the field. The functions $\{f_{\omega lm}\}$ and $\{g_{\omega lm}\}$ represent the ingoing positive and negative helicity modes, respectively. Their asymptotic forms on $\mathscr{I}^-$ are given by
\begin{equation}
\begin{aligned}
    f_{\omega lm} & \sim \frac{1}{\sqrt{8 \pi^2}} \frac{1}{r \sqrt{ \sin\theta}} e^{-i \omega v} e^{im\phi}  \varphi^+_{\omega l m}, \\
    g_{\omega lm} & \sim \frac{1}{\sqrt{8 \pi^2 }} \frac{1}{r \sqrt{ \sin\theta}} e^{+i \omega v} e^{-im\phi} \varphi^-_{\omega l m}.
\end{aligned}
\end{equation}
The angular dependence is encoded in the four-component spinors $\varphi^\pm_{\omega lm}$ as, 
\begin{align}
    \varphi^+_{\omega l m} = \begin{pmatrix}
    S_{1 \Lambda} (\theta) \\
    0  \\
    S_{1 \Lambda} (\theta) \\
    0
    \end{pmatrix}, \quad 
    \varphi^-_{\omega l m} = \begin{pmatrix}
    0 \\
    S_{2 (-\Lambda)} (\theta)  \\
    0 \\
    -S_{2 (-\Lambda)} (\theta)
    \end{pmatrix}.
\end{align}
Here, the spinor $\varphi^+$ corresponds to the positive helicity state ($L=+1$), while $\varphi^-$ corresponds to the negative helicity state ($L=-1$). 
The modes of negative helicity involve the set of quantum numbers $-\Lambda \equiv \{-\omega,l,-m\}$. The creation and annihilation operators for the ingoing modes on past null infinity, $\mathscr{I}^-$, are quantized by imposing the canonical anti-commutation relations, as was done in the Schwarzschild case shown in Eq.~\eqref{sch_fermion_ACR_negative}.

Similarly, the quantum field on $\mathscr{I}^+$ can be written interms of ingoing and outgoing modes as
\begin{equation}
    \hat{\Psi}=\int_0^{\infty} d\omega \sum_{l=s}^{\infty} \sum_{m=-l}^{+l} \left( p_{\omega lm}\hat{c}^-_{\omega lm}+q_{\omega lm}\hat{d}^{+}_{\omega lm}+r_{\omega lm}\hat{h}^-_{\omega lm}+s_{\omega lm}\hat{k}^{+}_{\omega lm} \right).
\end{equation}
In this expansion, $\{p_{\omega lm}\}$ and $\{q_{\omega lm}\}$ are the outgoing mode functions corresponding to positive and negative helicity modes detected by distant observers, while $\{r_{\omega lm}\}$ and $\{s_{\omega lm}\}$ represent the ingoing modes that are absorbed by the BH. The asymptotic forms of the outgoing modes are given in terms of the retarded time coordinate, $u$ as
\begin{equation}
\begin{aligned}
    p_{\omega lm} & \sim \frac{1}{\sqrt{8 \pi^2}} \frac{1}{r \sqrt{ \sin\theta}} e^{-i \omega u} e^{im\phi}  \varphi^+_{\omega l m}, \\
    q_{\omega lm} & \sim \frac{1}{\sqrt{8 \pi^2}} \frac{1}{r \sqrt{ \sin\theta}} e^{i \omega u} e^{-im\phi}  \varphi^-_{\omega l m}.
\end{aligned}
\end{equation}
All sets of operators are quantized by imposing the standard anti-commutation relations. The operators for the outgoing modes on $\mathscr{I}^+$, $\{\hat{c}, \hat{d}\}$, satisfy relations identical to those given in Eq.~\eqref{sch_fermion_ACR_positive}, and the ingoing operators, $\{\hat{h}, \hat{k}\}$, obey an analogous set.

\subsubsection*{Particle Creation for Fermionic modes}

Now to calculate the particle spectrum, we follow the same procedure by propagating the outging mode $p_{\omega lm}$ backward in time from future null infinity ($\mathscr{I}^+$) along the null geodesic $\gamma$ to past null infinity ($\mathscr{I}^-$). After scattering off the spacetime geometry, this mode (now denoted as $p^{(2)}_{\omega lm}$), has the asymptotic form on $\mathscr{I}^-$ as
\begin{equation}
    p_{\omega l m}^{(2)}  \sim \frac{1}{\sqrt{8 \pi^2}} \frac{1}{r \sqrt{ \sin\theta}} e^{-i\tilde{\omega} u(v) }e^{i m \tilde{\phi}} \varphi^+_{\omega l m}.
\end{equation}
Here, $\tilde{\phi} = \phi-\Omega_h t_0$ is the azimuthal coordinate in a frame co-rotating with the BH's event horizon far outside the collapsing body at some early times, and $\tilde{\omega} = \omega - m\Omega_h$ is the mode frequency measured in this frame. The relation between the retarded time $u$ and advanced time $v$ for the Kerr geometry is given in Eq.~\eqref{u_v_relation_kerr}.
The scattered mode is a superposition of the ingoing basis modes, described by a Bogoliubov transformation. The coefficients of this transformation are determined by the inner product on $\mathscr{I}^-$ as
\bea\label{bogo_kerr_fermion}
\begin{aligned}
    \alpha_{\omega lm \, \omega^{\prime}l^{\prime}m^{\prime}} &= (p_{\omega lm}^{(2)}, f_{\omega'l'm'})_{\mathscr{I}^-} \approx \frac{e^{i(m \Omega_h t_0-\omega^\prime v_0)}}{2\pi}  K^{i\frac{\tilde{\omega}}{\kappa}}  \left( -i \omega^\prime \right)^{\left(-1+i\frac{\tilde{\omega}}{\kappa}\right)}   \Gamma\left(1-i\frac{\tilde{\omega}}{\kappa}\right)  \delta_{ll^{\prime}}\delta_{mm^{\prime}}, \\    
    \beta_{\omega lm \, \omega^{\prime}l^{\prime}m^{\prime}} &= -(p_{\omega lm}^{(2)}, f^*_{\omega'l'm'})_{\mathscr{I}^-} \approx - \alpha_{\omega l m (-\omega^\prime) l^\prime m^\prime}.
\end{aligned}
\eea
The number of created particles in a given mode is found by integrating the squared modulus of the $\beta$ coefficient. This yields the total number of particles observed at $\mathscr{I}^+$ in the mode $(\omega, l, m)$ when the initial state was the ingoing vacuum $|0_-\rangle$
\begin{equation}
    \langle 0_- | \hat{c}_{\omega lm}^{+(2)} \hat{c}_{\omega lm}^{-(2)} | 0_- \rangle = \int_0^{\infty} d\omega^{\prime} \sum_{l^{\prime},m^{\prime}} |\beta_{\omega lm, \omega^{\prime}l^{\prime}m^{\prime}}|^2=\frac{\Gamma_{\omega lm}}{e^{\frac{2\pi \tilde{\omega} }{\kappa}}+1}\delta(0).
\end{equation}
The number of particles emitted per unit time and per unit frequency is found by summing over all angular modes as
\begin{equation}
    n_{\omega} = \sum_{l,m}\frac{\Gamma_{\omega lm}}{e^{(\omega-m\Omega_h)/T_{\rm BH}}+1},
\end{equation}
where $T_{\rm BH} = \kappa/2\pi$ is the Hawking temperature and $\Gamma_{\omega lm}$ is the graybody factor for the specific mode.

In the following section, we provide a brief overview of the Thermofield Dynamics (TFD) formalism, which we will later use to incorporate finite-temperature effects into the Hawking radiation spectrum.


\section{Preview of Thermofield Dynamics}\label{TFD}
To analyze the evolution of a thermal quantum field in a BH background, we employ the formalism of TFD, which introduces a thermal vacuum state to represent thermal averages as expectation values in a (twin) doubled Hilbert space~\cite{Das_book}. Consider a system in thermal equilibrium at a temperature $ T_b $. The ensemble average of an operator $ \hat{A} $ is given by
\begin{equation}
\langle \hat{A} \rangle_\beta = Z^{-1}(\beta) \text{Tr}(e^{-\beta \hat{H}} \hat{A}),
\end{equation}
where $ \beta = 1/T_b $, $ \hat{H} $ is the Hamiltonian of the system, and $ Z(\beta) = \text{Tr}(e^{-\beta \hat{H}}) $ is the partition function. Let $ \{ |n\rangle \} $ be the energy eigenstates of $ \hat{H} $ with eigenvalues $ E_n $, such that $ \hat{H} |n\rangle = E_n |n\rangle $ and $ \langle n | m \rangle = \delta_{nm} $, then one can write,
\begin{equation}
\langle \hat{A} \rangle_\beta = Z^{-1}(\beta) \sum_n e^{-\beta E_n} \langle n | \hat{A} | n \rangle.
\end{equation}
To express this thermal average as an expectation value in a vacuum-like state, we define the thermal vacuum $ |0, \beta \rangle $ in the following obvious way:
\begin{equation}\label{thermal_expectation}
\langle \hat{A} \rangle_\beta = \langle 0, \beta | \hat{A} | 0, \beta \rangle = Z^{-1}(\beta) \sum_n e^{-\beta E_n} \langle n | \hat{A} | n \rangle. 
\end{equation}
Assuming a linear decomposition in terms of the basis states, $|0, \beta \rangle = \sum_n f_n(\beta) |n\rangle,$ we find that
\begin{equation}\label{thermal_expansion}
\langle 0, \beta | \hat{A} | 0, \beta \rangle = \sum_{n,m} f_n^*(\beta) f_m(\beta) \langle n | \hat{A} | m \rangle. 
\end{equation}
Comparing Eqs.~\eqref{thermal_expansion} and \eqref{thermal_expectation}, we get
\begin{equation}
f_n^*(\beta) f_m(\beta) = Z^{-1}(\beta) e^{-\beta E_n} \delta_{nm}.
\end{equation}
This equation cannot be satisfied by complex numbers $ f_n $ alone. This is resolved by doubling the degree of freedom of the standard Hilbert space $\mathscr{H}$ by introducing a fictious system $\tilde{\mathscr{H}}$ an identical copy of the original system (also called tilde system) to form the Hilbert space $\mathscr{H} \otimes \tilde{\mathscr{H}}$~\cite{Das_book}. The new basis states are $ |n, \tilde{m}\rangle = |n\rangle \otimes |\tilde{m}\rangle $, where $ |\tilde{n}\rangle $ are eigenstates of the tilde system. The thermal vacuum is then expressed as
\bea
\vert 0, \beta \rangle =\sum_n f_n(\beta) \vert n,\tilde{n}\rangle = \sum_n f_n(\beta) \vert n\rangle\otimes \vert \tilde{n}\rangle .
\eea
The expectation value of an operator $ \hat{A} $ (acting on the original system) becomes
\bea
\langle 0, \beta | \hat{A} | 0, \beta \rangle = \sum_{n,m} f_n^*(\beta) f_m(\beta) \langle n, \tilde{n} | \hat{A} | m, \tilde{m} \rangle = \sum_n |f_n(\beta)|^2 \langle n | \hat{A} | n \rangle,
\eea
using the orthogonality $ \langle \tilde{n} | \tilde{m} \rangle = \delta_{nm} $ and the fact that the operators of one space do not acts on the basis of the other space. Therefore, the consistency with Eq.~\eqref{thermal_expectation} requires
\begin{equation}
|f_n(\beta)|^2 = Z^{-1}(\beta) e^{-\beta E_n},
\end{equation}
implying
\begin{equation}\label{eq:f_n}
f_n(\beta) = Z^{-1/2}(\beta) e^{-\beta E_n/2},
\end{equation}
i.e., the coefficients are real and temperature dependent. Note that while Eq.~\eqref{eq:f_n} could mathematically admit a complex phase for $f_n(\beta)$, such a phase does not contribute to the physical observables or the thermal averages in the TFD formalism~\cite{Das_book}. Therefore, without loss of generality, we adopt the standard convention of taking these coefficients to be real. This construction allows one to compute thermal expectation values as vacuum expectation values in the doubled Hilbert space using the thermal vacuum. Therefore the thermal state $\vert 0,\beta \rangle$ can be written in the product space basis $\vert n_{\omega lm},\tilde{n}_{\omega lm}\rangle$ as
\bea
\vert 0, \beta \rangle =Z^{-\frac{1}{2}}(\beta)\int_{\omega}d\omega\sum_{n_{\omega lm}} e^{-n_{\omega lm}\frac{\beta\omega}{2}} \vert n_{\omega lm},\tilde{n}_{\omega lm}\rangle.
\eea
Now we will construct a unitary operator $U(\theta)$ which will transform the product space vacuum $\vert 0, \tilde{0}\rangle$ to the thermal vacuum $\vert 0, \beta \rangle$ as
\beq{}
U(\theta) \vert 0, \tilde{0}\rangle=\vert 0, \beta \rangle .
\eeq
The form of this transformation depends on the statistics of the field.

This brings us to the most interesting conclusion of this section. The upshot is the following: Given any operator $\hat{A}$ acting on the appropriate Hilbert space defined above, it can be transformed into its thermal counterpart, denoted by $\hat{A}(\beta)$, through the unitary relation
\begin{equation}
    \hat{A}(\beta) = \hat{U}(\beta)\,\hat{A}\,\hat{U}^\dagger(\beta).
\end{equation}
The crucial property of this formalism is that the expectation value of any physical observable $\hat{\mathcal{O}}$ in the thermal state at inverse temperature $\beta$ is given by the expectation value in the TFD ground state: $\text{Tr}(\hat{\rho} \hat{\mathcal{O}}) = \langle 0, \tilde{0} \vert \hat{\mathcal{O}}(\beta) \vert 0, \tilde{0} \rangle$.

\subsubsection*{Bosonic Fields:}
For a bosonic field, the unitary transformation is generated by
\begin{equation}
\hat{U}_B(\beta) = \exp\left[ -\int d\omega \sum_{l,m} \theta_{\omega}(\beta) \left( \tilde{\hat{B}}^-_{\omega lm} \hat{B}^-_{\omega lm} - \hat{B}^+_{\omega lm} \tilde{\hat{B}}^+_{\omega lm} \right) \right],
\end{equation}
where the angle $\theta_{\omega}(\beta)$ is a function of temperature, defined by the relations
\begin{equation}\label{theta_boson}
    \cosh\theta_{\omega}(\beta) = \frac{1}{\sqrt{1-e^{-\beta\omega}}}, \qquad \sinh\theta_{\omega}(\beta) = \frac{e^{-\beta\omega/2}}{\sqrt{1-e^{-\beta\omega}}}.
\end{equation}
This operator transforms the original annihilation and creation operators $\hat{B}_{\omega lm}^{\pm}$ as
\begin{equation}
    \hat{B}^{\pm}_{\omega lm}(\beta) = \hat{U}_B(\beta)\,\hat{B}_{\omega lm}^{\pm} \, \hat{U}^{\dagger}_B(\beta) = \hat{B}_{\omega lm}^{\pm}  \cosh\theta_{\omega}(\beta) - \tilde{\hat{B}}^{\mp}_{\omega lm}\sinh\theta_{\omega}(\beta).
\end{equation}
The expectation value of the thermal number operator $\hat{N}_{\omega lm}(\beta) = \hat{B}^+_{\omega lm}(\beta)\hat{B}^-_{\omega lm}(\beta)$ in the ground state $\ket{0, \tilde{0}}$ correctly reproduces the Bose-Einstein distribution
\begin{equation}
    \langle 0, \tilde{0} \vert \hat{N}_{\omega lm}(\beta) \vert 0, \tilde{0} \rangle  = \frac{1}{e^{\beta\omega}-1} \quad  \forall \; l, m.
\end{equation}

\subsubsection*{Fermionic Fields:}
For a fermionic field, the Bogoliubov transformation is generated by
\begin{equation}
    \hat{U}_F(\beta) = \exp\left[- \int d\omega \sum_{l,m} \theta_{\omega}(\beta) \left(\tilde{\hat{C}}^-_{\omega lm}\hat{C}^-_{\omega lm}-\hat{C}^+_{\omega lm}\tilde{\hat{C}}^+_{\omega lm}  \right) \right].
\end{equation}
The angle $\theta_\omega(\beta)$ is now defined by trigonometric functions
\begin{equation}\label{theta_fermion}
    \cos\theta_{\omega}(\beta) = \frac{1}{\sqrt{1+e^{-\beta\omega}}}, \qquad \sin\theta_{\omega}(\beta) = \frac{e^{-\beta\omega/2}}{\sqrt{1+e^{-\beta\omega}}}.
\end{equation}
The transformation for the fermionic operators $\hat{C}^{\pm}_{\omega lm}$ is now given by
\begin{equation}
    \hat{C}^{\pm}_{\omega lm}(\beta) = \hat{U}_F(\beta)\,\hat{C}_{\omega lm}^{\pm} \, \hat{U}^{\dagger}_F(\beta) = \hat{C}_{\omega lm}^{\pm}  \cos\theta_{\omega}(\beta) - \tilde{\hat{C}}^{\mp}_{\omega lm}\sin\theta_{\omega}(\beta).
\end{equation}
This construction yields the correct Fermi-Dirac distribution for the particle number in the TFD ground state as
\begin{equation}
    \langle 0, \tilde{0} \vert \hat{N}_{\omega lm}(\beta) \vert 0, \tilde{0} \rangle = \frac{1}{e^{\beta\omega}+1} \quad  \forall \; l, m.
\end{equation}
The field operators in the thermal state, $\hat{\Phi}(\beta)$ and $\hat{\Psi}(\beta)$, are then obtained by replacing the original operators in their respective mode decompositions with these thermally transformed operators.

\section{Black Hole in a Thermal Bath}\label{Finite_Hawking}

In realistic cosmological settings, such as the early Universe, BHs are not isolated but are typically immersed in a thermal environment. This surrounding bath of particles can significantly alter their evaporative properties through the process of stimulated emission. To incorporate these effects, we adopt the TFD formalism, which we have discussed in the previous section. 

Consider a massless scalar and fermionic field propagating in a BH background, immersed in a thermal bath at a fixed temperature $T_b = 1/\beta$. To account for the bath, the quantum fields must be described using a thermal representation~\cite{Kalita:2025foa}. The key physical observable is the flux of particles radiated to future null infinity, $\mathscr{I}^+$. This is calculated by taking the expectation value of the number operator for outgoing modes, $\hat{N}^+_{\omega lm}(\beta)$, in the state $\ket{0,\tilde{0}}_-$ that is defined on past null infinity, $\mathscr{I}^-$. The connection between the past and future is established through the Bogoliubov transformation that relates the ingoing and outgoing field modes. A detail calculation (see Appendix~\ref{tfd_hawking} for details) shows that the particle number for a given mode $(\omega, l, m)$ is given by 
\begin{equation}
    {}_-\langle 0,\tilde{0} \vert \hat{N}^+_{\omega lm}(\beta) \vert 0,\tilde{0} \rangle_- = \int_0^{\infty}d\omega^{\prime} \sum_{l^{\prime},m^{\prime}} \left[\, f_1(\theta_\omega) |\beta_{\omega lm, \omega'l'm'}|^2 + f_2(\theta_\omega) |\alpha_{\omega lm, \omega'l'm'}|^2 \,\right],
    \label{general_spectrum}
\end{equation}
where $\alpha_{\omega lm, \omega'l'm'}$ and $\beta_{\omega lm, \omega'l'm'}$ are the Bogoliubov coefficients for the spacetime, while the functions $f_{1}(\theta_\omega),$ and $f_{2}(\theta_\omega)$ depend on the thermal angle $\theta_\omega$ and the spin/statistics of the field.


\subsection{Altered Thermal spectrum for Scalar Fields in BH background}

For a bosonic scalar field, the thermal transformation is described by hyperbolic functions, where $f_1(\theta_\omega) = \cosh^2\theta_\omega(\beta)$ and $f_2(\theta_\omega) = \sinh^2\theta_\omega(\beta)$. Applying this to Eq.~\eqref{general_spectrum} with the appropriate Bogoliubov coefficients for the Schwarzschild geometry \eqref{bogo_sch_boson}, we can easily derive the total particle flux per unit frequency
\begin{equation}
    n_{\omega} = \frac{\Gamma_{\omega}}{e^{2\pi\omega/\kappa} - 1} \left[1 + \frac{e^{2\pi\omega/\kappa} + 1}{e^{\beta \omega} - 1} \right].
\end{equation}
This expression demonstrates the combined effect of spontaneous Hawking emission and stimulated emission induced by the thermal environment. The modification of the thermal spectrum is obviously over and above the standard black body spectrum of black holes, and indeed, if $\beta\rightarrow\infty$, we shall recover the standard Hawking spectrum, while if no black holes are present, the Boltzmann spectrum is obtained.

To extend our analysis to the case of a rotating Kerr BH, due to the presence of rotation, the frequency $\omega$ is effectively replaced by the combination $(\omega - m \Omega_h)$, as can be seen in the expressions for the Bogoliubov coefficients Eqs.~\eqref{bogo_kerr_boson}. Consequently, the number density of particles observed at future null infinity becomes
\begin{equation}\label{Kerr_corrected_spectrum_boson}
    n_{\omega} = \sum_{l,m} \frac{\Gamma_{\omega lm}}{e^{(\omega - m\Omega_h)/T_{\rm BH}} - 1} \left[1 + \frac{e^{(\omega - m\Omega_h)/T_{\rm BH}} + 1}{e^{\beta \omega} - 1} \right],
\end{equation}
where $T_{\rm BH}$ is the Hawking temperature of the Kerr BH. This result captures both the superradiance-induced frequency shift and the thermal amplification.

\subsection{Altered Thermal spectrum for Fermionic Fields in BH background}
For a fermionic field, the TFD transformation involves trigonometric functions, with $f_1(\theta_\omega) = \cos^2\theta_\omega(\beta)$ and $f_2(\theta_\omega) = \sin^2\theta_\omega(\beta)$. The resulting emission spectrum for a Schwarzschild BH, using the Bogoliubov coefficients from \eqref{bogo_sch_fermion}, is a modified Fermi-Dirac distribution
\begin{equation}
    n_{\omega} = \frac{\Gamma_{\omega}}{e^{2\pi\omega/\kappa}+1}\left[1+ \frac{e^{2\pi\omega/\kappa}-1}{e^{\beta\omega}+1} \right].
\end{equation}
This expression is extended to the rotating Kerr black holes by incorporating the frequency shift $\tilde{\omega} = \omega - m\Omega_h$ as in \eqref{bogo_kerr_fermion}, which gives the final particle flux
\begin{equation}\label{Kerr_corrected_spectrum_fermion}
    n_{\omega} = \sum_{l,m} \frac{\Gamma_{\omega lm}}{e^{(\omega-m\Omega_h)/T_{\rm BH}}+1}\left[1+ \frac{e^{(\omega-m\Omega_h)/T_{\rm BH}}-1}{e^{\beta\omega}+1} \right].
\end{equation}

The derived expressions for the Kerr BH, Eqs.~\eqref{Kerr_corrected_spectrum_boson} and \eqref{Kerr_corrected_spectrum_fermion}, are the general results for particle emission in a thermal bath. As expected, in the non-rotating limit where the specific angular momentum $a_* \to 0$ (and thus $\Omega_h \to 0$), they reduce smoothly to their Schwarzschild counterparts. Therefore, in the main body of this work, we will use these general Kerr expressions to calculate the BH mass and spin decay rates, specializing to the Schwarzschild case by setting the rotation parameter to zero where appropriate.

\section{Decaying black holes}\label{BH_decay}

In addition to particle emission, Hawking radiation leads to a gradual loss of mass and angular momentum from the BH over time. In this section, we derive the evolution equations for the BH mass $M(t)$ and spin $J(t)$ using the finite temperature corrected spectrum obtained earlier. We begin by expressing the number spectrum for particles of the $i^{\text{th}}$ species in the presence of a thermal bath. For a field with spin $s_i$, the number density at energy $E_i$ takes the form
\begin{equation}
n_{E_i} = \sum_{l=s_i}\sum_{m=-l}^{l} \frac{\Gamma_{\omega lm}(s_i)}{e^{\frac{E_i - m \Omega_h}{T_{\rm BH}}} - (-1)^{2 s_i}} \left[1 + \frac{e^{\frac{E_i - m \Omega_h}{T_{\rm BH}}} + (-1)^{2 s_i}}{e^{\beta E_i} - (-1)^{2 s_i}}\right],
\end{equation}
where $\Gamma_{\omega lm}(s_i)$ is the greybody factor for spin-$s_i$ fields. The energy is related to the particle mass and momentum by $E_i^2 = \mu_i^2 + p^2$. To compute the net particle production due to the BH (excluding the contribution from the thermal bath alone), we subtract the purely thermal background part as
\bea
\tilde{n}_{E_i}=n_{E_i}-\sum_{l=s_i} \sum_{m=-l}^{l} \frac{\Gamma_{\omega lm}(s_i)}{e^{\beta E_i}-1}=\sum_{l=s_i} \sum_{m=-l}^{l}\frac{\Gamma_{\omega lm}(s_i)}{e^{\frac{E_i-m\Omega_h }{T_{\rm BH}}}-(-1)^{2s_i}}\left(1+2 \frac{(-1)^{2s_i}}{e^{\beta E_i}-(-1)^{2s_i}} \right).\, \;\;\;\;
\eea
\begin{figure}[t]
\centering
\includegraphics[scale=1]{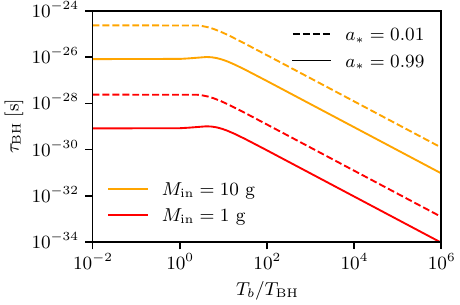} 
\caption{Lifetime of BHs vs. bath temperature $T_b$ (Eq.~\eqref{lifetime}) for two initial masses, $1\,\mathrm{g}$ (red) and $10\,\mathrm{g}$ (yellow), shown for Schwarzschild (dashed) and Kerr (solid) BHs.}
\label{Lifetime_vs_T_b}
\end{figure}
The particle emission rate per energy interval is then given by~\cite{Page:1976df}
\begin{equation}
\frac{d^2 N_i}{dE_i dt} = \frac{g_i}{2\pi} \tilde{n}_{E_i},
\end{equation}
where $g_i$ denotes the number of degrees of freedom of the $i^{\text{th}}$ species. Switching from energy $E_i$ to momentum $p$ using $E_i dE_i = p dp$, we obtain the emission rate per momentum interval as
\begin{equation}
\frac{d^2 N_i}{dp dt}=g_i \sum_{l=s_i} \sum_{m=-l}^{l} \frac{d^2 N_i^{lm}}{dp dt},
\end{equation}
where the particle production rate of each mode is given by
\begin{equation}
\frac{d^2 N_i^{lm}}{dp \, dt}=\frac{1}{2\pi} \,\left[\frac{\Gamma_{\omega lm}(s_i)}{e^{\frac{E_i-m\Omega_h }{T_{\rm BH}}}-(-1)^{2s_i}}\right]\left[1+2 \frac{(-1)^{2s_i}}{e^{\beta E_i}-(-1)^{2s_i}} \right] \frac{p}{E_i}.
\end{equation}

In the high-energy limit ($GMp \gg 1$), the greybody factor becomes approximately independent of spin and approaches the geometric optics limit: $\Gamma_{\omega lm}(s_i) \approx 27 G^2 M^2 p^2$~\cite{Cheek:2021odj}. We define the ratio
\begin{equation}
\varphi_{\omega lm}(s_i) = \frac{\Gamma_{\omega lm}(s_i)}{27 G^2 M^2 p^2},
\end{equation}
which encodes the deviation from the geometric optics approximation. With this the total mass loss rate due to Hawking radiation can now be computed by integrating the energy flux of all emitted particles and summing over all species as~\cite{Page:1976df}
\begin{equation}\label{BH_mass_evolution_eq}
\frac{dM}{dt} = -\sum_i \int_0^\infty E_i \frac{d^2 N_i}{dp dt} dp = -\epsilon \frac{M_p^4}{M^2},
\end{equation}
where $M_p = 1/\sqrt{G} \approx 1.22 \times 10^{19}$ GeV is the Planck mass. To the leading order approximation, the above equation can be integrated as
\begin{equation}
M \simeq M_{\rm in}[1-\Gamma_{_{\rm BH}} (t-t_{\rm in })]^{\frac 1 3},
\end{equation}
with $M_{\rm in}$ being the initial mass of the BH at time $t_{\rm in}$ and the BH decay life time can be written from the above equation as,
\begin{equation}\label{lifetime}
\tau_{_{\rm BH}} = \frac {1}{\Gamma_{_{\rm BH}}} \simeq \frac{M_{\rm in}^3}{3 \epsilon M_p^4} .
\end{equation}
\begin{figure}[t]
\centering
\includegraphics[width=.49\textwidth]{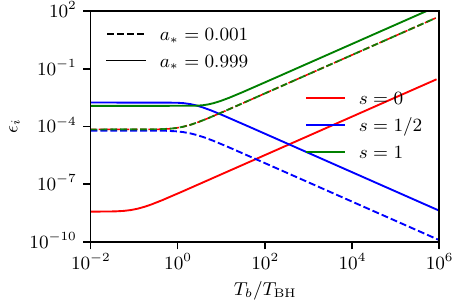}
\includegraphics[width=.49\textwidth]{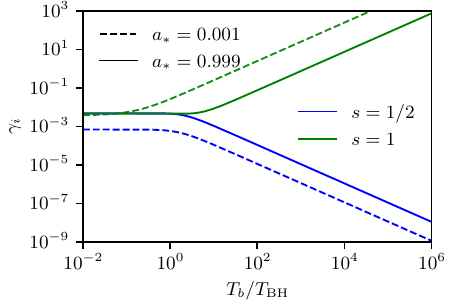}
\caption{\textbf{Left Panel:} Variation of $\epsilon_i$ (Eq.~\eqref{epsilon_i}) with bath temperature $T_b$ for Schwarzschild (dashed lines) and Kerr (solid lines) BHs, plotted for three particle spins: scalar ($s = 0$, red), fermion ($s = 1/2$, blue), and vector boson ($s = 1$, green). \textbf{Right Panel:} Dependence of $\gamma_i$ (Eq.~\eqref{gamma_i}) on $T_b$ for particles with spin $s = 1/2$ (blue) and $s = 1$ (green), shown for both Schwarzschild (dashed) and Kerr (solid) BHs.}
\label{fig:ep_gam}
\end{figure}
%
%
Fig.~\ref{Lifetime_vs_T_b} shows the BH lifetime ($\tau_{\rm BH}$) versus bath temperature ($T_b$) for two initial configurations of $1\,\mathrm{g}$ (red) and $10\,\mathrm{g}$ (yellow), for  Schwarzschild (dashed) and for Kerr (solid) BHs. As expected, on increasing bath temperature, the lifetime of the BHs decreases. The dimensionless efficiency factor $\epsilon=\sum_i g_i \epsilon_i$ contains the thermal correction contributions and
\begin{align}\label{epsilon_i}
\epsilon_i &= \frac{27}{8192 \pi^5} \int_{z_i}^{\infty} dx \sum_{l=s_i} \sum_{m=-l}^{l} \frac{\varphi_{\omega lm}(s_i) (x^2 - z_i^2)}{e^{\frac{x'}{2 f(a_)}} - (-1)^{2 s_i}} \left(1 + 2 \frac{(-1)^{2 s_i}}{e^{x / T'_b} - (-1)^{2 s_i}}\right) x,
\end{align}
where we introduce dimensionless variables for numerical convenience $x=8 \pi GM E_i $, $z_i=8 \pi GM \mu_i$,  $T_b'=T_b/T_{\rm BH}$, $x' = x-8 \pi GMm \Omega_h$ and
\begin{equation}
f(a_*)=\frac{\sqrt{1-a_*^2}}{1+\sqrt{1-a_*^2}} .
\end{equation}
The left panel of Fig.~\ref{fig:ep_gam} shows $\epsilon_i$ as a function of bath temperature $T_b$ for Schwarzschild (dashed) and Kerr (solid) BHs, considering three particle spins: $s=0$ (red), $s=1/2$ (blue), and $s=1$ (green). In the high bath temperature limit, $T_b^{\prime} \rightarrow \infty$, $\epsilon_i \propto T_b^{\prime}$ for bosonic particles, while for fermion $\epsilon_i \rightarrow 0$.


\begin{figure}[t]
\centering
\includegraphics[width=.49\textwidth]{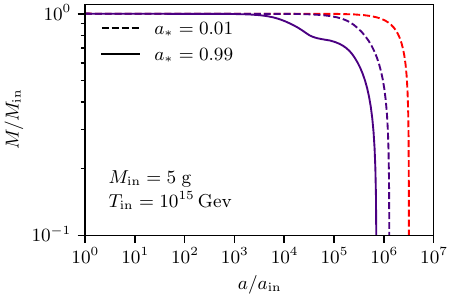}
\hspace{-0.2cm}
\includegraphics[width=.49\textwidth]{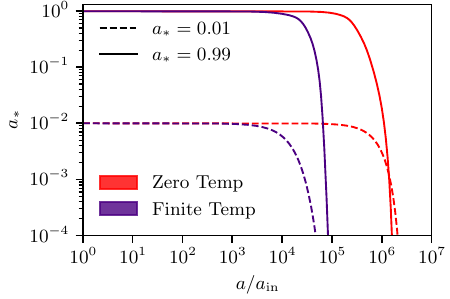}
\caption{\textbf{Left Panel:} Evolution of BH mass as a function of the scale factor for an initial mass of $5\, \mathrm{g}$ and initial bath temperature $10^{15}\, \mathrm{GeV}$. The solid purple line corresponds to the Kerr BH with thermal corrections, the dashed purple line to the Schwarzschild BH with thermal corrections, and the red dashed line to the Schwarzschild BH without thermal corrections. \textbf{Right Panel:} Evolution of the spin parameter $a_*$ for the same initial conditions as the left panel.}
\label{fig:mass_spin}
\end{figure}

Similar to the BH mass decay rate as each particle carries off angular momentum $m$ about the axis of the hole, the angular momentum of the hole $J $ decreases as~\cite{Cheek:2021odj, Page:1976df, Auffinger:2022khh, Arbey:2019jmj}
\begin{equation}\label{dJ_dt}
\frac{dJ}{dt} = -\sum_i \int_0^{\infty} \sum_{l=s_i} \sum_{m=-l}^{l} m \frac{d^2 N_i^{lm}}{dp dt} dp = -a_{\ast} \gamma  \frac{M_p^2}{M}
\end{equation}
with $\gamma=\sum_i \gamma_i$ is the angular momentum evaporation function and
\begin{equation}\label{gamma_i}
\gamma_i= \frac{27 }{1024 \pi^4}  \int_{z_i}^{\infty} \sum_{l=s_i} \sum_{m=-l}^{l} \frac{m a_{\ast}^{-1} \varphi_{\omega lm}(s_i) (x^2-z_i^2) }{e^{\frac{x^{\prime}}{2f(a_{\ast})}}- (-1)^{2 s_i}} \left(1+2 \frac{(-1)^{2s_i}}{e^{\frac{x}{T^{\prime}_b}}-(-1)^{2s_i}} \right) dx .
\end{equation}
Substituting the defination of $J$ ($\equiv a_* G M^2$) into Eq.~\eqref{dJ_dt}, one finds the evolution equation of the spin parameter as
\begin{equation}\label{BH_spin_evolution_eq}
\frac{da_{\ast}}{dt}=- a_{\ast} \left(\gamma - 2\epsilon \right) \frac{M_p^4}{M^3} 
\end{equation}
The right panel of Fig.~\ref{fig:ep_gam} displays the variation of $\gamma_i$ with bath temperature $T_b$ for spins $s = 1/2$ (blue) and $s = 1$ (green), comparing Schwarzschild (dashed) and Kerr (solid) BHs. Similar to $\epsilon_i$, $\gamma_i$ increases with $T_b$ for bosons and decreases for fermions. The right panel of Fig.~\ref{fig:mass_spin} shows the evolution of the spin parameter $a_*$ as a function of the scale factor $a(t)$ for two initial values, $a_*=0.01$ and $a_*=0.99$, under both zero-temperature and finite-temperature conditions. The qualitative difference between the scalar and fermionic emission rates in the presence of a thermal bath can be understood through the statistical properties of the fields. For bosons, the thermal background leads to Bose enhancement, where the outgoing flux is amplified. Conversely, for fermions, the Pauli exclusion principle leads to Pauli blocking, where the occupation of final states by the thermal bath inhibits further emission. This result demonstrates that the TFD formalism correctly captures the quantum statistical back-reaction of the environment on the BH evaporation process.

This analysis allows us to track how the BH gradually transitions towards the non-rotating limit and eventually evaporates completely. In the subsequent section, we apply this formalism to model-independent reheating scenarios and analyze its impact on the evaporation lifetime of PBHs.

\section{PBH during reheating}\label{Reheating}

The cosmological reheating phase provides a compelling setting to investigate finite temperature corrections to Hawking radiation, particularly due to the presence of a dynamically evolving thermal background. To explore these effects, we consider a generic reheating scenario in which the temperature of the ambient radiation decreases more slowly than the standard $T \propto a^{-1}$ scaling, owing to continuous entropy injection from the decay of the inflaton field.

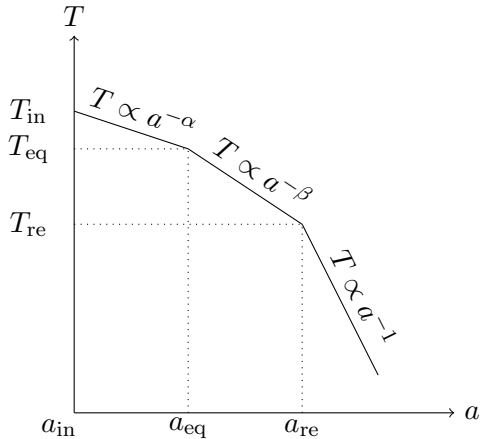
\begin{figure}[t]
    \centering
    \begin{tikzpicture}[scale=1]
        \draw[->] (0, 0) -- (5, 0) node[right] {$a$};
        
        \draw[->] (0, 0) -- (0, 5) node[above] {$T$};
        
        \node at (-0.2, -0.2) {$a_{\rm in}$};
		\node at (1.5, -0.2) {$a_{\rm eq}$};
		\node at (3, -0.2) {$a_{\rm re}$};        
        
        \node at (-0.6, 4) {$T_{\rm in}$};
        \node at (-0.6, 3.5) {$T_{\rm eq}$};
        \node at (-0.6, 2.5) {$T_{\rm re}$};

        \draw[black] (0, 4) -- (1.5, 3.5) node[above,sloped,pos=0.55] {$T \propto a^{-\alpha}$};
        \draw[black] (1.5, 3.5) -- (3, 2.5) node[above,sloped,pos=0.55]{$T \propto a^{-\beta}$};
        \draw[black] (3, 2.5) -- (4, 0.5) node[above,sloped,pos=0.55]{$T \propto a^{-1}$};
        
		\draw[dotted] (0, 3.5) -- (1.5, 3.5);
		\draw[dotted] (0, 2.5) -- (3, 2.5); 
		
		\draw[dotted] (1.5, 0) -- (1.5, 3.5);
		\draw[dotted] (3, 0) -- (3, 2.5);        
        
    \end{tikzpicture}
    \caption{Schematic evolution of background radiation temperature.}
    \label{temp_evolution}
\end{figure}
We adopt a parametrized temperature evolution, illustrated in Fig.~\ref{temp_evolution}, in which the background radiation temperature starts from a maximum value $T_{\rm in}$ immediately after the end of inflation. It subsequently scales as $T \propto a^{-\alpha}$ until it decreases to an intermediate temperature $T_{\rm eq}$. Beyond this epoch, the scaling changes to $T \propto a^{-\beta}$ and continues until the reheating temperature $T_{\rm re}$ is reached. At $T_{\rm re}$, the energy densities of radiation and the inflaton field become equal: $\rho_{\phi} = \rho_R = \pi^2 g_{\ast}(T) T_{\rm re}^4/30$. For our analysis, we assume $g_{\ast}(T) \simeq 106.75$, corresponding to the Standard Model particle content~\cite{Husdal:2016haj}. After $T_{\rm re}$, the Universe transitions to the standard radiation-dominated era, with the temperature scaling as $T \propto a^{-1}$. The transition temperature $T_{\rm eq}$ can be expressed in terms of $\alpha$, $\beta$, and other background parameters as  
\begin{equation}  
T_{\rm eq} = \left[\left( \frac{\pi^2}{30} g_{\ast}(T) \right)^{-1} \rho^{\rm in} T_{\rm in}^{-\frac{3}{\alpha}(1+\omega)} T_{\rm re}^{\frac{3}{\beta}(1+\omega)-4} \right]^{\frac{\alpha\beta}{3(1+\omega)(\alpha - \beta)}},  
\end{equation}  
where $\rho^{\rm in}$ is the background energy density at $T_{\rm in}$, and $\omega$ denotes the equation of state. 
Consequently, $\rho^{\rm in}$ can be written as $\rho^{\rm in} = \rho^{\rm end} \exp \left\{ 3(1+\omega) \right\}$.
For our numerical analysis, we adopt a benchmark value for the energy density at the end of inflation of $\rho_{\text{end}} \sim 10^{63} \, \text{GeV}^4$, which corresponds to the current observational upper bound on the scale of inflation~\cite{Planck:2018jri, BICEP:2021xfz}.


Additionally, the two temperature scaling regimes, $T \propto a^{-\alpha}$ and $T \propto a^{-\beta}$ correspond to distinct phases of energy injection and thermalization. For example, in models where the inflaton has multiple decay channels, such as decays into bosons $\phi \rightarrow ss$ (or $\phi\phi \rightarrow ss$), the background radiation temperature evolves as $T \propto a^{-3(1-\omega)/8}$ (or $T \propto a^{-9(1-\omega)/8}$) for non-gravitational decays. On the other hand, if the inflaton decays into fermions, $\phi \rightarrow \bar{f}f$, the temperature scales as $T \propto a^{-3(1+\omega)/8}$~\cite{Haque:2023yra}. Appropriately choosing the background equations of state $\omega$, one can in principle get different scaling of radiation temperature during reheating phases. In this work we adopt $\alpha = 0.16$ and $\beta = 0.98$ as representative values that facilitate a prolonged interaction between the PBH and the thermal bath, allowing for a clear assessment of the lifetime reduction. Further discussions on different reheating mechanisms and their underlying particle-production dynamics can be found in~\cite{Haque:2023awl, RiajulHaque:2023cqe, Meyers:2013gua, Asadi:2019iod, Nguyen:2019kbm, Amin:2014eta}.

\begin{figure}[t]
\centering
\includegraphics[width=.49\textwidth]{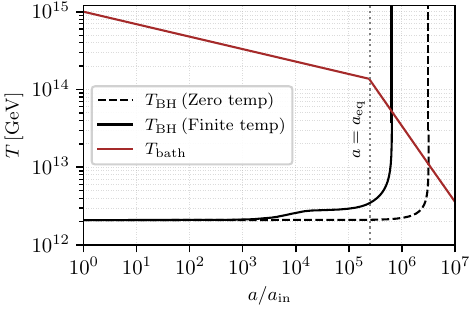}
\hspace{-0.2cm}
\includegraphics[width=.49\textwidth]{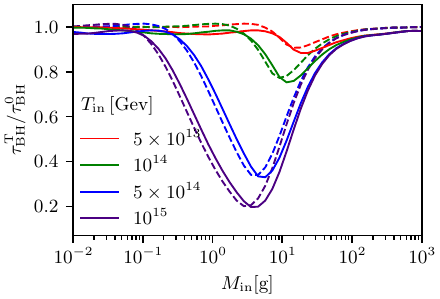}
\caption{\textbf{Left Panel:} Evolution of the BH temperature $T_{\rm BH}$ and the ambient bath temperature $T_{\rm bath}$ as functions of the scale factor $a/a_{\rm in}$. The dashed black curve denotes the zero-temperature Schwarzschild BH temperature, while the solid black curve corresponds to the finite-temperature Kerr BH case for an initial mass $M_{\rm in} = 5 \rm g$ and initial bath temperature $T_{\rm in} = 10^{15}\rm Gev$. The brown curve shows the background bath temperature. The vertical dotted line marks the point $a=a_{\rm eq}$. \textbf{Right Panel:} The ratio of PBH lifetimes at finite temperature to those at zero temperature, $\tau_{\rm BH}^{\rm T}/\tau_{\rm BH}^{0}$, is shown as a function of the initial PBH mass $M_{\rm in}$ for both Schwarzschild (dashed) and Kerr (solid) BHs, across different initial background temperatures $T_{\rm in}$.}
\label{fig:T_tau}
\end{figure}

In the following, we assume pressureless background energy density  ($\omega= 0$), which prolongs the duration of the reheating phase. This allows PBHs to persist longer in the regime of non-trivial temperature evolution, emphasizing the effects of the evolving bath temperature on PBH dynamics. To capture this influence, we numerically solve the coupled equations governing the evolution of PBH mass and spin [Eqs.~\eqref{BH_mass_evolution_eq} and \eqref{BH_spin_evolution_eq}] for various initial spin values. Throughout this work, we assume that all PBHs form at the initial radiation temperature $T_{\rm in}$, treating the initial PBH mass $M_{\rm in}$ as a free parameter. While the standard scenario of PBH formation via the collapse of overdensities fixes the initial mass to be of order the Hubble horizon mass ($M_{\rm PBH} \propto M_H$), we consider a broader class of physically well-motivated formation mechanisms. For instance, PBHs can form during phase transitions at the GUT or PeV scale~\cite{Anantua:2008am, Zagorac:2019ekv, Liu:2021svg, Escriva:2022yaf}, leading to a broad spectrum, or via quantum nucleation processes at the end of inflation (the Bousso--Hawking mechanism~\cite{Bousso:1996au, BoussoHawking1999}), where the initial mass is not strictly tied to the horizon size.

To illustrate the effect of a finite thermal environment on the BH temperature, left panel of Fig.~\ref{fig:T_tau} shows the evolution of $T_{\rm BH}$ together with the bath temperature $T_{\rm bath}$ as functions of the scale factor for a reheating temperature of $T_{\rm re} = 10^{-2} \, \text{GeV}$. At early times the finite-temperature correction is negligible, and the two $T_{\rm BH}$ curves coincide. As the Universe expands and the bath temperature drops, the finite-temperature BH solution begins to deviate from its zero-temperature counterpart. The left panel of Fig.~\ref{fig:mass_spin} shows the evolution of PBH mass as a function of the scale factor for an initial mass of $5 \, \text{g}$ and $T_{\rm in} = 10^{15} \, \text{GeV}$. Three cases are compared: a Kerr PBH with finite temperature corrections (solid purple), a Schwarzschild PBH with finite-temperature corrections (dashed purple), and a Schwarzschild PBH in a zero temperature background (red dashed).


The corresponding PBH lifetime $\tau_{\rm BH}$ is evaluated using the relation $a(t) \propto t^{2/3(1+\omega)}$, giving
\begin{equation}
\tau_{\rm BH} = t_{\rm in} \left[ \left( \frac{a_{\rm ev}}{a_{\rm in}} \right)^{2} - 1 \right],
\end{equation}
where $t_{\rm in}$ is the formation time and $a_{\rm ev}$ is the scale factor at evaporation. The right panel of Fig.~\ref{fig:T_tau} shows the ratio of PBH lifetimes at finite temperature $\tau_{\rm BH}^{\rm T}$ to those at zero temperature $\tau_{\rm BH}^{\rm 0}$ as a function of the initial PBH mass $M_{\rm in}$ both for Schwarzschild (dashed) and Kerr (solid) PBHs for different $T_{\rm in}$ values.

The plot reveals that low-mass PBHs exhibit similar lifetimes regardless of thermal corrections, as their Hawking temperature eventually exceeds the background temperature, making thermal effects negligible. As PBH mass increases, the finite temperature effect also weakens. This is because a significant portion of particle emission from a PBH occurs near the end of its lifetime. For higher masses, PBHs decay later when the background temperature has significantly decreased. By the time of decay, the PBH temperature surpasses the background temperature, thereby reducing the impact of finite temperature effects. Interestingly, increasing $T_{\rm in}$ amplifies the finite temperature effect. Higher $T_{\rm in}$ values result in an elevated background temperature near the final moments of PBH evaporation, enhancing the interplay between the PBH temperature and the evolving background.

For efficient (perturbative/non-perturbative) reheating, the maximum temperature is typically of order $T_{\rm in} \sim 10^{15}\,\mathrm{GeV}$, whereas for gravitational reheating it is expected to be lower, $T_{\rm in} \sim 10^{12}\,\mathrm{GeV}$. The range of maximum temperatures explored in our analysis therefore covers a broad class of reheating scenarios, rendering our results widely applicable and largely model independent.

It is worth noting that our analysis assumes a spatially uniform temperature for the ambient thermal bath. While local energy deposition from the PBH could, in principle, create `hotspots' or temperature gradients, such effects are expected to be suppressed in the high-temperature, high-density environment of the early Universe. Specifically, the formation of hotspots via the Landau-Pomeranchuk-Migdal (LPM) effect~\cite{He:2022wwy, Hamaide:2023ayu} is most efficient when $T_{\rm BH} \gg T_{\rm bath}$, whereas our formalism focuses on the interplay where $T_{\rm bath}$ is high enough to significantly modify the evaporation rate. Even if local thermal fluctuations were present, they would likely lead to a further enhancement of the evaporation rate, making our current findings a conservative estimate of the PBH lifetime reduction.

\section{Conclusion}\label{conclusion}

In this paper, we have looked into the effect of thermal environment on the Hawking radiation emitted from black holes, focusing particularly on Kerr BHs immersed in a cosmological thermal bath. In such a setting, the emitted particles interact with the thermal background and thermalize, leading to modifications in the Hawking radiation spectrum. This is quite expected since thermal fluctuations now accompanies quantum fluctuations in this framework. To extract these effects in an unified way, we employ the formalism of finite temperature quantum field theory, and in particular the theory of thermofield dynamics (TFD), to derive the corrected occupation number spectrum, which in turn alters the rates of loss of mass and spin of the progenitor BH. This approach provides a consistent framework at the level of field operators for treating quantum fields in thermal backgrounds, thus extending the standard vacuum-based Hawking radiation framework.

Applying this formalism to PBHs formed in the early Universe, we have explored their evolution in a model independent reheating background, where the thermal bath evolves non trivially due to entropy injection from the decaying inflaton field. We solve the coupled mass and spin evolution equations for PBHs in this thermal setting and demonstrate that the presence of a thermal bath leads to an enhanced emission rate. Our results show that the PBH lifetime is reduced by approximately an order of magnitude for a maximum initial temperature of $10^{15} \, \rm GeV$. Notably, the effect is slightly more pronounced for Kerr BHs compared to their Schwarzschild counterparts. Therefore, the basic ideas presented here may be considered to form the basis for further developments into this field of quantum field theory in curved spacetimes.

These results underscore the importance of including thermal effects when analyzing PBH evaporation in realistic early Universe scenarios. Our formalism opens a new avenue to study BH thermodynamics beyond the isolated vacuum approximation, with potential implications for cosmology, gravitational wave backgrounds, and dark matter constraints.


\acknowledgments

DM wishes to acknowledge support from the Science and Engineering Research Board (SERB), Department of Science and Technology (DST), Government of India (GoI), through the Core Research Grant CRG/2020/003664. AC thanks the DAE- BRNS for their funding through 58/14/25/2019-BRNS. The work of JK is supported by the Ministry of Human Resource Development, Government of India.


\appendix

\section{Dirac Fields in Black Hole Spacetimes}
\label{appendix_dirac_formalism}

This appendix provides the necessary components for understanding the behaviour of Dirac spinors in both the Schwarzschild and the Kerr spacetimes. We outline the method by first recalling the metric, tetrads, gamma matrices, and spin connection for each geometry. 

Note that since
$GL(4, \mathbb{R})$ does not have a finite dimenisional double valued representation, one requires that the fields be mapped to the internal  tangent space at each point where the metric is Minkowskian. Therefore, describing a fermionic field $\Psi$ in a curved spacetime with metric $g_{\mu\nu}$ requires the introduction of a local inertial frame at each point. This is achieved through the tetrad formalism, which connects the curved spacetime to the flat tangent space via the tetrad fields $e^a_\mu$ and their inverses $e^\mu_a$, satisfying $g_{\mu\nu} = e^a_\mu e^b_\nu \eta_{ab}$. The flat-space metric signature is taken to be $\eta_{ab} = \text{diag}(-1, 1, 1, 1)$, and the flat-space gamma matrices, $\gamma^a$, are in the chiral representation:
\begin{equation}
    \gamma^0 = 
    \begin{pmatrix}
    iI_2 & 0 \\
    0 & -iI_2
    \end{pmatrix},
    \quad
    \gamma^j = 
    \begin{pmatrix}
    0 & i\sigma_j \\
    -i\sigma_j & 0
    \end{pmatrix}
\end{equation}
with $I_2$ is $2\times2$ identity matrix and
\begin{equation}
\sigma_1=\begin{pmatrix}
0 &1 \\
1 & 0
\end{pmatrix},\;\;\sigma_2=\begin{pmatrix}
0 & -i\\
i & 0
\end{pmatrix},\;\;\sigma_3=\begin{pmatrix}
1 & 0 \\
0 & -1
\end{pmatrix}
\end{equation}
are the Pauli spin matrices. These gamma matrices satisfy the Clifford algebra $\{\gamma^a, \gamma^b\} = 2\eta^{ab}I_4$. The curved-space gamma matrices are constructed via the inverse tetrad, $\gamma^\mu = e^\mu_a \gamma^a$, and the spin connection is given by $\Omega_\mu = \frac{1}{8}\omega_{ab\mu}[\gamma^a, \gamma^b]$ and $\omega_{ab\, \mu} = \eta_{ac}e^c_{\nu}\left[\partial_{\mu}e^{\nu}_b+ e^{\sigma}_b\Gamma^{\nu}_{\sigma\mu} \right]$.
\subsubsection*{Schwarzschild geometry}
For the static and spherically symmetric Schwarzschild spacetime, it is convenient to work in tortoise coordinates $(t, r^*, \theta, \phi)$, where the line element is
\begin{equation}
    ds^2 = f(r)(-dt^2 + dr^{*2}) + r^2(d\theta^2 + \sin^2\theta d\phi^2),
\end{equation}
with $f(r) = 1-r_g/r$ and $r_g=2GM$. The diagonal nature of this metric allows for a simple diagonal tetrad. The non-zero covariant components are~\cite{Muller:2009bw}
\begin{align}
    e^0_t = f^{1/2}, \qquad e^1_{r^*} = f^{1/2}, \qquad e^2_\theta = r, \qquad e^3_\phi = r\sin\theta ,
\end{align}
whereas the non-zero contravariant (inverse) components are
\begin{align}
    e^t_0 = f^{-1/2}, \qquad e^{r^*}_1 = f^{-1/2}, \qquad e^\theta_2 = \frac{1}{r}, \qquad e^\phi_3 = \frac{1}{r\sin\theta}.
\end{align}
From these tetrads, the non-zero spin connection coefficients $\omega_{ab\mu} = -\omega_{ba\mu}$ are found to be
\begin{equation}
\begin{aligned}
    \omega_{01t} = -\frac{f'}{2f} = -\frac{r_g}{2r^2} , \quad
    \omega_{12\theta} = -f^{1/2} , \quad
    \omega_{13\phi} = -\sin\theta f^{1/2} , \quad
    \omega_{23\phi} = -\cos\theta 
\end{aligned}
\end{equation}
where $f^\prime = df/dr^*$. This yields the spin connection components $\Omega_\mu$ to be
\begin{equation}
\begin{aligned}
    \Omega_t &= -\frac{1}{4}\frac{f'}{f}[\gamma^0, \gamma^1], & \Omega_{r^*} &= 0, \\
    \Omega_\theta &= -\frac{1}{4}f^{1/2}[\gamma^1, \gamma^2], &
    \Omega_\phi &= -\frac{1}{4}\sin\theta f^{1/2}[\gamma^1, \gamma^3] - \frac{1}{4}\cos\theta[\gamma^2, \gamma^3].
\end{aligned}
\end{equation}
%

\subsubsection*{Kerr Spacetime}
For the stationary and axisymmetric Kerr spacetime, we use the Boyer-Lindquist coordinates $(t,r,\theta,\phi)$. The line element is
\begin{equation}
    ds^2 = -\frac{\Delta}{\Sigma^2}\left[dt - a_k\sin^2\theta d\phi\right]^2 + \frac{\Sigma^2}{\Delta}dr^2 + \Sigma^2 d\theta^2 + \frac{\sin^2\theta}{\Sigma^2}\left[(r^2+a_k^2)d\phi - a_k dt\right]^2 ,
\end{equation}
where $r_g = 2GM$, $a_k = J/M$, $\Sigma^2 = r^2 + a_k^2\cos^2\theta$, and $\Delta = r^2 - r_g r + a_k^2$. The spacetime's rotation requires a non-diagonal tetrad basis. The chosen basis one-forms $e^a = e^a_\mu dx^\mu$ are
\begin{equation}
\begin{aligned}
    e^0 &= \sqrt{\frac{\Delta}{\Sigma^2}}dt - \sqrt{\frac{\Delta}{\Sigma^2}}a_k\sin^2\theta d\phi, & e^1 &= \sqrt{\Sigma^2}d\theta, \\
    e^2 &= \frac{\sin\theta}{\sqrt{\Sigma^2}}(r^2+a_k^2)d\phi - \frac{\sin\theta}{\sqrt{\Sigma^2}}a_k dt, & e^3 &= \sqrt{\frac{\Sigma^2}{\Delta}}dr .
\end{aligned}
\end{equation}
The corresponding dual basis vectors $e_a = e^\mu_a \partial_\mu$ are
\begin{equation}
\begin{aligned}
    e_0 &= \frac{(r^2+a_k^2)}{\sqrt{\Delta\Sigma^2}}\partial_t + \frac{a_k}{\sqrt{\Delta\Sigma^2}}\partial_\phi, & e_1 &= \frac{1}{\sqrt{\Sigma^2}}\partial_\theta ,\\
    e_2 &= \frac{a_k\sin\theta}{\sqrt{\Sigma^2}}\partial_t + \frac{1}{\sin\theta\sqrt{\Sigma^2}}\partial_\phi ,& e_3 &= \sqrt{\frac{\Delta}{\Sigma^2}}\partial_r .
\end{aligned}
\end{equation}
The corresponding curved-space gamma matrices are constructed from these tetrads to be
\begin{equation}
\begin{aligned}
    \gamma^t &= \frac{(r^2+a_k^2)}{\sqrt{\Delta\Sigma^2}}\gamma^0 + \frac{a_k\sin\theta}{\sqrt{\Sigma^2}}\gamma^2, \qquad \gamma^r = \sqrt{\frac{\Delta}{\Sigma^2}}\gamma^3 , \\
    \gamma^\theta &= \frac{1}{\sqrt{\Sigma^2}}\gamma^1 , \qquad \gamma^\phi = \frac{a_k}{\sqrt{\Delta\Sigma^2}}\gamma^0 + \frac{1}{\sin\theta\sqrt{\Sigma^2}}\gamma^2 .
\end{aligned}
\end{equation}
The resulting components of the spin connection $\Omega_\mu$ are
\begin{equation}
\begin{aligned}
    \Omega_t &= -\frac{r_g(r^2-a_k^2\cos^2\theta)}{4\Sigma^4}\gamma^0\gamma^3 + \frac{rr_ga_k\cos\theta}{2\Sigma^4}\gamma^1\gamma^2, \\
    \Omega_r &= \frac{ra_k\sin\theta}{2\sqrt{\Delta\Sigma^2}}\gamma^0\gamma^2 + \frac{a_k^2\sin\theta\cos\theta}{2\sqrt{\Delta\Sigma^2}}\gamma^1\gamma^3, \\
    \Omega_\theta &= -\frac{\sqrt{\Delta}a_k\cos\theta}{2\Sigma^2}\gamma^0\gamma^2 + \frac{r\sqrt{\Delta}}{2\Sigma^2}\gamma^1\gamma^3, \\
    \Omega_\phi &= \frac{\sqrt{\Delta}a_k\sin\theta\cos\theta}{2\Sigma^2}\gamma^0\gamma^1 + \frac{a_k\sin^2\theta}{4\Sigma^4}\left(\Sigma^2(2r-r_g)+2r_gr^2\right)\gamma^0\gamma^3 \\
    & \quad -\frac{A\cos\theta}{2\Sigma^4}\gamma^1\gamma^2 + \frac{\sqrt{\Delta}r\sin\theta}{2\Sigma^2}\gamma^2\gamma^3,
\end{aligned}
\end{equation}
where $A=(r^2+a_k^2)^2-\Delta \, a_k^2 \sin^2 \theta$.


\section{Hawking Radiation in a Thermal Bath via TFD}
\label{tfd_hawking}
We consider massless scalar and fermionic fields propagating in a BH background, which is assumed to be in equilibrium with a thermal bath at temperature $T_b = 1/\beta$. To account for the bath, the quantum fields are described using the Thermo-Field Dynamics (TFD) formalism, wherein the standard creation and annihilation operators are replaced by their thermal counterparts~\cite{Kalita:2025foa}.

\subsection*{Scalar Fields}
At past null infinity, $\mathscr{I}^-$, the scalar field in the thermal state is expanded in terms of the ingoing modes $\{ f_{\omega lm} \}$
\begin{equation}
    \hat{\Phi}(\beta) = \int_0^\infty d\omega \sum_{l,m} \left( f_{\omega lm} \hat{a}_{\omega lm}^-(\beta) + f^*_{\omega lm} \hat{a}_{\omega lm}^+(\beta) \right),
\end{equation}
where the thermal annihilation and creation operators are given by the Bogoliubov transformation
\begin{equation}
    \hat{a}^{\pm}_{\omega lm}(\beta) = \hat{a}_{\omega lm}^{\pm}  \cosh\theta_{\omega}(\beta)-\tilde{\hat{a}}^{\mp}_{\omega lm}\sinh\theta_{\omega}(\beta),
\end{equation}
with the thermal mixing angle $\theta_\omega(\beta)$ defined by the Bose-Einstein distribution, as given in \eqref{theta_boson}. The number operator in the TFD ground state, $\ket{0,\tilde{0}}_- = \ket{0_-} \otimes \ket{\tilde{0}_-}$, correctly yields the thermal spectrum of the bath
\begin{equation}
    {}_-\bra{0,\tilde{0}} \hat{a}^+_{\omega lm}(\beta) \hat{a}^-_{\omega lm}(\beta) \ket{0,\tilde{0}}_- = \frac{1}{e^{\beta\omega} - 1} \quad \forall \; l,m.
\end{equation}
This confirms that the thermal operators properly encode the presence of the background thermal environment.

Analogously, the scalar field at future null infinity, $\mathscr{I}^+$, is expanded in terms of both outgoing ($p_{\omega lm}$) and ingoing ($q_{\omega lm}$) modes as
\begin{equation}
    \hat{\Phi}(\beta)=\int_0^{\infty} d\omega \sum_{l,m} \left[p_{\omega lm}\hat{b}^-_{\omega lm}(\beta)+p^{*}_{\omega lm}\hat{b}^+_{\omega lm}(\beta)+q_{\omega lm}\hat{c}^-_{\omega lm}(\beta)+q^{*}_{\omega lm}\hat{c}^+_{\omega lm}(\beta) \right].
\end{equation}
The key physical observable is the spectrum of outgoing particles at $\mathscr{I}^+$, as seen from the perspective of the past vacuum state $\ket{0,\tilde{0}}_-$. This expectation value reflects both the standard Hawking radiation and stimulated emission from the thermal bath
\begin{equation}
    {}_-\bra{0,\tilde{0}} \hat{b}^+_{\omega lm}(\beta)\hat{b}^-_{\omega lm}(\beta)\ket{0,\tilde{0}}_- = \int_0^{\infty}d\omega^{\prime} \sum_{l^{\prime},m^{\prime}} \left[ \cosh^2 \theta_{\omega}(\beta) |\beta_{\omega lm, \omega'l'm'}|^2 + \sinh^2 \theta_{\omega}(\beta) |\alpha_{\omega lm, \omega'l'm'}|^2 \,\,\right].
\end{equation}
%

\subsection*{Fermionic Fields}
The analysis for a massless fermionic field in the thermal bath proceeds in a similar fashion, with differences arising due to Fermi-Dirac statistics. The field on past null infinity, $\mathscr{I}^-$, is decomposed as
\begin{equation}
    \hat{\Psi}(\beta)=\int_0^{\infty} d\omega \sum_{l,m} \left[f_{\omega lm}\hat{a}^-_{\omega lm}(\beta)+g_{\omega lm}\hat{b}^+_{\omega lm}(\beta) \right],
\end{equation}
where the thermal operators for particles ($\hat{a}^{\pm}_{\omega lm}(\beta)$) and anti-particles ($\hat{b}^{\pm}_{\omega lm}(\beta)$) are given by the fermionic Bogoliubov transformation
\begin{equation}
\begin{aligned}
    \hat{a}^{\pm}_{\omega lm}(\beta) &= \hat{a}_{\omega lm}^{\pm}  \cos\theta_{\omega}(\beta)-\tilde{\hat{a}}^{\mp}_{\omega lm}\sin\theta_{\omega}(\beta), \\
    \hat{b}^{\pm}_{\omega lm}(\beta) &= \hat{b}_{\omega lm}^{\pm}  \cos\theta_{\omega}(\beta)-\tilde{\hat{b}}^{\mp}_{\omega lm}\sin\theta_{\omega}(\beta),
\end{aligned}
\end{equation}
with the angle $\theta_{\omega}(\beta)$ defined by the Fermi-Dirac distribution in \eqref{theta_fermion}. The field on future null infinity, $\mathscr{I}^+$, is likewise expanded in terms of thermal operators for its outgoing and ingoing modes
\begin{equation}
    \hat{\Psi}(\beta)=\int_0^{\infty} d\omega \sum_{l,m}\left( p_{\omega lm}\hat{c}^-_{\omega lm}(\beta)+q_{\omega lm}\hat{d}^+_{\omega lm}(\beta)+r_{\omega lm}\hat{h}^-_{\omega lm}(\beta)+s_{\omega lm}\hat{k}^+_{\omega lm}(\beta) \right).
\end{equation}
The outgoing particle spectrum, as measured in the ingoing vacuum $\ket{0,\tilde{0}}_-$, is found by calculating the expectation value of the outgoing number operator, $\hat{c}^+_{\omega lm}(\beta)\hat{c}^-_{\omega lm}(\beta)$. This yields the Hawking radiation corrected by the thermal bath
\begin{equation}
    {}_-\bra{0,\tilde{0}} \hat{c}^+_{\omega lm}(\beta)\hat{c}^-_{\omega lm}(\beta)\ket{0,\tilde{0}}_- = \int_0^{\infty}d\omega^{\prime} \sum_{l^{\prime},m^{\prime}} \left( \cos^2 \theta_{\omega}(\beta) |\beta_{\omega lm, \omega'l'm'}|^2 + \sin^2 \theta_{\omega}(\beta) |\alpha_{\omega lm, \omega'l'm'}|^2 \right).
\end{equation}


\end{document}